\documentclass[preprint]{ptephy}

\preprintnumber{%
KEK-TH-245
}

\usepackage{amssymb}
\usepackage{amsthm}
\usepackage{amsmath}
\usepackage{booktabs}
\usepackage{bbm}
\usepackage{mathtools}
\usepackage{subcaption}
\usepackage{braket}
\usepackage{here}
\usepackage{hyperref}

\newcommand{\Slash}[1]{{\ooalign{\hfil/\hfil\crcr$#1$}}}
\newcommand{\be}{\begin{equation}}
\newcommand{\ee}{\end{equation}}
\newcommand{\lt}{\left}
\newcommand{\rt}{\right}
\newcommand{\del}{\partial}
\newcommand{\non}{\nonumber \\}
\newcommand{\fn}{\footnote}

\numberwithin{equation}{section}

\usepackage{url}

\DeclareMathOperator*{\SumInt}{%
\mathchoice%
  {\ooalign{$\displaystyle\sum$\cr\hidewidth$\displaystyle\int$\hidewidth\cr}}
  {\ooalign{\raisebox{.14\height}{\scalebox{.7}{$\textstyle\sum$}}\cr\hidewidth$\textstyle\int$\hidewidth\cr}}
  {\ooalign{\raisebox{.2\height}{\scalebox{.6}{$\scriptstyle\sum$}}\cr$\scriptstyle\int$\cr}}
  {\ooalign{\raisebox{.2\height}{\scalebox{.6}{$\scriptstyle\sum$}}\cr$\scriptstyle\int$\cr}}
}

\begin{document}

\title{QED on the lattice and numerical perturbative computation of $g-2$}

\author{%
\name{\fname{Ryuichiro} \surname{Kitano}}{1,2}
\name{\fname{Hiromasa} \surname{Takaura}}{1,\ast}
}

\address{%
\affil{1}{KEK Theory Center, Tsukuba 305-0801, Japan}
\affil{2}{Graduate University for Advanced Studies (Sokendai), Tsukuba 305-0801, Japan}
\email{hiromasa.takaura@yukawa.kyoto-u.ac.jp}
}

\date{\today}

\begin{abstract}
We compute the electron $g$ factor to the $\mathcal{O}(\alpha^5)$ order
on the lattice in quenched QED.
We first study finite volume corrections in various IR regularization 
methods to discuss which regularization is optimal 
for our purpose.
We find that in QED$_L$ the finite volume correction to the
effective mass can have different parametric dependences
depending on the size of Euclidean time $t$
and match the `naive on-shell result' only at very large $t$ region, $t \gg L$. 
We adopt finite photon mass
regularization to suppress
finite volume effects exponentially 
and also discuss 
our strategy for selecting simulation parameters and
the order of extrapolations to
efficiently obtain the $g$ factor.
We perform lattice simulation using small lattices
to test feasibility of our calculation strategy.
This study can be regarded as an intermediate step toward giving  the five-loop coefficient independently of the preceding studies. 
\end{abstract}

\subjectindex{B01, B38, B50, B59}

\maketitle

\section{Introduction}
The anomalous magnetic moment of the electron, the electron $g-2$,
is one of the most precisely measured observables in 
particle physics. 
It thus gives us a good opportunity to test our understanding of quantum field theory.
Theoretically the perturbative calculation of QED contributions has reached the five-loop [$\mathcal{O}(\alpha^5)$] level \cite{Aoyama:2012wk,Aoyama:2017uqe},
although a slight discrepancy in the five-loop coefficient is reported \cite{Volkov:2019phy}.
The five-loop contribution is indeed relevant to the precision of the experimental measurements \cite{2008,Fan:2022eto}.
In order to examine the consistency of the SM prediction with the experimental result,
the value of the fine structure constant is crucial, yet the value is not determined consistently among different experiments \cite{Morel:2020dww}.

In ref.~\cite{Kitano:2021ecc}, we proposed a new method to calculate the perturbative series of the electron $g-2$ using 
numerical stochastic perturbation theory (NSPT) \cite{DiRenzo:1994av,DiRenzo:1994sy,DiRenzo:2000qe,DiRenzo:2002vg,DiRenzo:2004hhl}.
We calculated the perturbative series to the three-loop level on the small lattices.
This was the first attempt to apply NSPT to QED observables. 
Since it can provide us with a new and alternative approach to 
calculating the perturbative series  to high orders and can have a wide range of application,
it is worth testing the usefulness of the method further.

In order to perform meaningful 
numerical simulations, 
we need to understand finite volume (FV)
corrections, which are known to be  
severe in lattice QED because of massless photon. 
In the first half of this paper, we
study FV corrections in various ways of IR regularization such as
subtractions of zero modes (known as QED$_L$ or QED$_{TL}$) and finite photon
mass~\cite{Hayakawa:2008an, Davoudi:2018qpl, Endres:2015gda} to
understand what kind of IR regularization is optimal, although FV
corrections are not understood enough in our previous
work~\cite{Kitano:2021ecc}. 
QED$_L$ is a famous and well-adopted regularization, and
FV corrections in this 
regularization have been studied in many papers 
such as refs.~\cite{Hayakawa:2008an,Davoudi:2014qua,Borsanyi:2014jba,Lubicz:2016xro,Davoudi:2018qpl,Bijnens:2019ejw,DiCarlo:2021apt}. 
Mainly based on results in ref.~\cite{Davoudi:2018qpl}, we add a new insight into 
FV correction in QED$_L$.
We find that 
the FV correction to the effective mass can have different
parametric dependences depending on the size of Euclidean time;
at $1/m \ll t \ll L$ the FV correction is given by $\mathcal{O}(t/ ( m L^2))$ while
at $t \gg L$ it is given by $\mathcal{O}(1/(mL))$.
The latter case matches the `naive on-shell result'
but this is not always true for general $t$.
From the discussions in this part we conclude that the massive photon regularization is the most controlled method for our computation.

In the second half of this paper, we perform lattice simulation of  
the electron $g-2$ in quenched QED, i.e., QED without the dynamical
electron. We adopt finite photon mass regularization, which is found to be most suited for our purpose. In quenched QED,
i.e., in sub-diagrams without lepton loops, there is a discrepancy in
the five-loop perturbative coefficient between
refs.~\cite{Aoyama:2017uqe, Aoyama:2019ryr} and
ref.~\cite{Volkov:2019phy}. Our study can potentially give an
independent result. As an attempt, we perform a five-loop level
calculation on the lattice. Our present study does not quite give a
conclusive result due to small lattice sizes. We regard the study in
this paper as an intermediate step toward obtaining the continuum
limit result of the five-loop coefficient.

The achievements of the present paper can be stated as follows.
First, the higher order calculation than our previous study \cite{Kitano:2021ecc}
is made possible.
This is because we use a method to
suppress backward propagations,
which are a
serious obstacle in the analysis in
ref.~\cite{Kitano:2021ecc}.
The higher order calculation can be
done also because numerical costs
to generate configurations are
drastically reduced due to 
quenched QED. In quenched QED, 
interaction terms are absent and 
the Langevin equation 
becomes trivial.
Therefore, configurations are
generated
according to a Gaussian distribution \cite{Luscher:2014mka, DallaBrida:2017tru, DallaBrida:2017pex}.
In this sense, the present paper
tests efficiency
of numerical calculation of
perturbative series on the lattice
rather than NSPT itself.
Secondly, we discuss in detail the
strategy for selecting simulation parameters and also the order of various extrapolations.
This is based on understanding of
systematic errors such as 
FV corrections, finite photon mass
effects, which are studied in this 
paper.

The paper is organized as follows. In Sec.~\ref{sec:2},
we study FV corrections in various 
IR regularization method to 
discuss what kind of IR
regularization we should adopt.
We also add a new insight into finite volume 
corrections in QED$_L$.
In Sec.~\ref{sec:3}, we perform a lattice simulation.
We first explain the outline of our calculation, and then we
study systematic uncertainties of
our calculation  to
discuss the strategy for selecting
simulation parameters and 
the order of various extrapolations.
Then we perform our numerical simulation following the strategy
to examine its feasibility. 
Sec.~\ref{sec:4} is devoted to the conclusions and discussion.


\section{Finite volume corrections in lattice QED}
\label{sec:2}

Before we attempt to calculate any physical quantities in QED, it is essential to provide a concrete definition of QED.
Specially, in a finite volume, the treatment of infrared divergence requires careful consideration to ensure that predictions agree with QED in continuum and infinite volume spacetime as a certain limit.
We below discuss and compare various definitions proposed or used in the literature. We will see that the regularization by finite photon mass is the most controlled method for our $g-2$ computations.

In Sec.~\ref{sec:2.1}, we study
FV corrections in QED$_L$ and QED$_{TL}$
to a momentum-space correlator at one loop.
This part includes already known facts 
and can be regarded as a review part.
In Sec.~\ref{sec:2.2},
we study FV corrections in QED$_L$ to a Euclidean 
time correlator, i.e., Fourier transform of
the momentum-space correlator, using a result in Sec.~\ref{sec:2.1}.
We point out that a FV effect
has different parametric dependences depending on
the size of $t/L$. 
This is the new and main result in this section.
In Sec.~\ref{sec:2.3}, 
we consider massive photon theory
and confirm exponential suppression of FV corrections for clarity.

\subsection{Finite volume corrections in QED$_L$ and QED$_{TL}$: momentum-space correlator
}
\label{sec:2.1}

We consider FV effects in various IR regularization methods: QED$_L$,
QED$_{TL}$,  and massive photon
regularization~\cite{Hayakawa:2008an, Davoudi:2018qpl,
Endres:2015gda}. The precise meaning of these regularizations is
explained shortly.

We consider the following quantity as an example:
\be
I=\SumInt  \frac{1}{k^2+m_{\gamma}^2} \frac{1}{(k+p)^2+m^2} . \label{consideredI}
\ee
Here $k$ denotes loop momentum  
and $p$ external momentum. The summation/integral symbol represents
the sum/integration of the loop momentum $k$, and its precise meaning
depends on regularization schemes as discussed below.
The above quantity mimics the one-loop correction
to the two-point function in scalar QED. 
$m_{\gamma}$ denotes photon mass, which will
be set to zero or non-zero below. In lattice QED, the zero mode of
loop momentum $k=(0,0,0,0)$ makes the result diverge. Therefore some
regularization of the zero mode is needed. One way is to introduce
non-zero photon mass. There are alternative methods which do not
introduce photon mass but modify the range of loop momentum sum.  
QED$_L$ and QED$_{TL}$ define $\SumInt$ as follows:
\begin{align}
&{\text{QED$_L$ on $\mathbb{T}^3 \times \mathbb{R}$}:} 
\qquad{}  \SumInt=\frac{1}{L^3} \sum_{\vec{k} \in B \mathbb{Z}_L^3\backslash \{0\} } \int \frac{d k_4}{2 \pi}  , \non 
&{\text{QED$_L$ on $\mathbb{T}^4$}:} 
\qquad{}  \SumInt=\frac{1}{L^3 T} \sum_{k_4 \in B \mathbb{Z}_T} \sum_{\vec{k} \in B \mathbb{Z}_L^3\backslash \{0\} } , \non
&{\text{QED$_{TL}$ on $\mathbb{T}^4$}:} 
\qquad{}  \SumInt=\frac{1}{L^3 T} \sum_{k \in B \mathbb{Z}^4_{TL} \backslash \{0\}}  ,
\end{align}
where $\vec{k}=(k_1,k_2,k_3)$, $B \mathbb{Z}_{L}=\frac{2 \pi}{L} \mathbb{Z}=\{\frac{2 \pi}{L} n | n \in \mathbb{Z} \}$, $B \mathbb{Z}_{T}=\frac{2 \pi}{T} \mathbb{Z}$, 
$B \mathbb{Z}^4_{TL} =B \mathbb{Z}_{L}^3 \times B \mathbb{Z}_{T}$, and $\backslash \{0\}$ means removal of the element $(0,\cdots,0)$.
In these regularization methods, 
the photon mass is set to zero.
We note that, although 
$\mathbb{T}^3 \times \mathbb{R}$ cannot be
realized in actual lattice simulations,
it is convenient to consider 
QED$_{L}$ on $\mathbb{T}^3 \times \mathbb{R}$ 
as an intermediate step for 
clarifying the difference between QED 
on $\mathbb{R}^4$ and an above theory, 
as  explained in ref.~\cite{Borsanyi:2014jba}.
In other words, we can clarify 
FV corrections of, for instance,
QED$_L$ on $\mathbb{T}^4$, 
considering the chain 
(QED on $\mathbb{R}^4$) $\to$ 
(QED$_{L}$ on $\mathbb{T}^3 \times \mathbb{R}$)
$\to$
(QED$_{L}$ on $\mathbb{T}^4$)
and studying the difference of each deformation.

Equation~\eqref{consideredI} mimics a two point function in scalar QED
and is not directly related to the $g$ factor.
Nevertheless, we understand some features of the regularization methods
and obtain implications by studying this simple integral.

Below we study FV corrections in QED$_L$ and QED$_{TL}$
setting $p=(0,0,0,p_4)$ with $p_4 \in \mathbb{R}$.
We note that the FV corrections to $I$ have been studied 
in the case of the on-shell momentum $p_4=im$ \cite{Borsanyi:2014jba, Davoudi:2018qpl}
and in the case of off-shell momentum $p_4 \in \mathbb{R}$ \cite{Davoudi:2018qpl}.
We give the FV corrections for the off-shell momentum also in this paper in a self-contained manner, 
as this result is used in the subsequent subsection.
The reason why we are interested in the off-shell momentum case rather than the on-shell momentum case is that
a Euclidean time correlator, which is what one actually obtains in lattice simulations, is equivalent to the Fourier integral of a momentum-space correlator $I$,
where the integral variable $p_4$ runs from $-\infty $ to $\infty$, namely {\it off-shell momenta}.
Although one might expect that this Fourier integral effectively sets $p_4$
to the on-shell value due to the residue theorem, we point out that
this understanding is too naive.  
This will be
discussed in Sec.~\ref{sec:2.2} after we review
FV corrections to a momentum-space correlator here. 

\vspace{2mm}
\noindent
{\bf{QED on $\mathbb{R}^4$ $\to$ 
QED$_{L}$ on $\mathbb{T}^3 \times \mathbb{R}$}} \\
First we consider the difference between QED$_L$ 
on  $\mathbb{T}^3 \times \mathbb{R}$ and 
QED on $\mathbb{R}^4$. It is 
given by \cite{Hayakawa:2008an}
\begin{align}
\Delta I(L) \equiv
&\lt(\SumInt_{\text{QED on $\mathbb{R}^4$}} -  \SumInt_{\text{QED$_L$ on  $\mathbb{T}^3 \times \mathbb{R}$}} \rt) 
\frac{1}{k^2} \frac{1}{(k+p)^2+m^2} \non
&=-\lt( \sum_{\vec{x} \in L \mathbb{Z}^3 \backslash \{0\} } -\frac{1}{L^3} \int d^3 x  \rt)
\int \frac{d^4 k}{(2\pi)^4} \frac{1}{k^2} \frac{1}{(k+p)^2+m^2} e^{i \vec{k} \cdot \vec{x}} \non
&=-\frac{1}{(4 \pi)^2} \int_0^1 dy \int_0^{\infty} ds \, s^{-1} e^{-\frac{s}{4 \pi} [y(1-y) p^2+y m^2] L^2} \lt(\vartheta_3 (0,e^{-\pi/s})^3-1-s^{3/2} \rt) . \label{DeltaI}
\end{align}
Here we used the Poisson resummation to rewrite the momentum sum $\sum_{\vec{k} \in B \mathbb{Z}_L^3 \backslash{0}}$ by 
the momentum integration $\int d^3 \vec{k}/(2 \pi)^3$ while introducing $\vec{x} \in L \mathbb{Z}$, where
$L \mathbb{Z}=\{L n | n \in \mathbb{Z}\}$.  
$ \int d^3 x $ corresponds to the subtraction of the spatial zero mode (since it gives $(2 \pi)^3 \delta^3(\vec{k})$)
and we rewrote the propagators by the Feynman parameter ($y$) integral and performed the momentum integration.
$\vartheta_3$ denotes the elliptic theta function, $\vartheta_3(0,e^{-\pi/s})=\sum_{n=-\infty}^{\infty} e^{-\pi n^2/s}$.

To study the asymptotic form of $\Delta I$ for $L \to \infty$, we consider
\be
\widetilde{\Delta I}(u) \equiv \int_0^{\infty} dL \, L^{-u-1} \Delta I (L). \label{Laplace}
\ee
We can see the asymptotic behavior of $\Delta I$ by studying singularities of this function.
If $\Delta I$ behaves as $\Delta I \sim L^{-u_0}$ for $L \gg 1$, 
$\widetilde{\Delta I}(u)$ develops 
singularities at $u=-u_0$ due to a divergence 
of the integral of eq.~\eqref{Laplace} 
around $L \sim  \infty$.

We obtain
\begin{align}
&\widetilde{\Delta I}(u) \non
&=-\frac{1}{32 \pi^2} \frac{1}{(4 \pi)^{u/2}}\Gamma(-u/2) 
\int_0^{1} dy \, [y(1-y) p^2+y m^2]^{u/2} \non
&\quad{}\times \int_0^{\infty} ds \, s^{u/2-1} \lt(\vartheta_3 (0,e^{-\pi/s})^3-1-s^{3/2} \rt) . \label{DeltaItilde}
\end{align}
The $y$-integral and the $s$-integral are factorized.
The first singularity of $\widetilde{\Delta I}(u)$ at negative $u$ is located at $u=-2$, where the $y$-integral diverges.
This tells us that the asymptotic behavior of $\Delta I(L)$ is $\Delta I (L) \sim L^{-2}$. The $s$-integral is convergent for $u=-2$.
To examine the convergence of the $s$-integral, it is convenient to keep the following relation in mind:
\begin{align}
\vartheta_3(0,e^{-\pi/s})
=\sum_{n=-\infty}^{\infty} e^{-\frac{\pi}{s} n^2}
=\sum_{m=-\infty}^{\infty} \int \frac{dk}{2 \pi} e^{-\frac{k^2}{4 \pi s}} e^{i k m}
=s^{1/2} \sum_{m=-\infty}^{\infty}  e^{-\pi s m^2}=s^{1/2} \vartheta_3(0,e^{-\pi s}) ,
\label{thetarelation}
\end{align}
where we used the Poisson resummation in the first equality and then performed the Gaussian integral. 
Thus one can see that
\begin{align}
&\vartheta_3 (0,e^{-\pi/s})^3-1 \sim (1+2 e^{-\pi/s})^3-1 & \text{for $s \ll 1$} \non
&\vartheta_3 (0,e^{-\pi/s})^3-s^{3/2} \sim  [s^{1/2} (1+2 e^{-s \pi})]^3-s^{3/2}  & \text{for $s \gg 1$} 
\end{align}
are very suppressed functions.
The expansion of $\widetilde{\Delta I}(u)$ around $u=-2$ is given by
\be
\widetilde{\Delta I}(u)=-\frac{\kappa_{-2}}{4 \pi}  \frac{1}{p^2+m^2} \frac{1}{u+2} +\mathcal{O}((u+2)^0) ,
\ee
where 
\be
\kappa_{-2} \equiv \int_0^{\infty} ds \, s^{-2}  \lt(\vartheta_3 (0,e^{-\pi/s})^3-1-s^{3/2} \rt) \simeq -2.837 .
\ee

The inverted formula of eq.~\eqref{Laplace} 
is given by
\be
\Delta I (L)=\frac{1}{2 \pi i} \int_{-i \infty-0}^{ i \infty-0} du \, L^{u} \widetilde{\Delta I}(u) . \label{inverse}
\ee
Calculating eq.~\eqref{inverse} by changing the integration path $-i \infty \to  i \infty$ to the contour surrounding the pole at $u=-2$,
we obtain
\be
\Delta I (p,L) =- \frac{\kappa_{-2}}{4 \pi} \frac{1}{p^2+m^2} \frac{1}{L^2}+(\text{higher order in $1/L$}) . \label{EuclidDeltaI}
\ee 
We note that {\it{the FV correction is singular at the on-shell momentum}}.
This result is not new and can be understood as a special case of eq.~(46) in ref.~\cite{Davoudi:2018qpl}.

\vspace{2mm}
\noindent
{\bf{QED$_{L}$ on $\mathbb{T}^3 \times \mathbb{R}$
$\to$ 
QED$_{L}$ on $\mathbb{T}^4$}} \\
Next we consider the difference between QED$_L$ on  $\mathbb{T}^3 \times \mathbb{R}$ and QED$_L$ on  $\mathbb{T}^4$:
\begin{align}
\Delta' I(L,T)
&=\lt(\SumInt_{\text{QED$_L$ on  $\mathbb{T}^3 \times \mathbb{R}$}} - \SumInt_{\text{QED$_L$ on  $\mathbb{T}^4$}}\rt) 
\frac{1}{k^2} \frac{1}{(k+p)^2+m^2} \non
&=-\frac{1}{L^3} \sum_{\vec{k} \in B \mathbb{Z}_L^3 \backslash \{0\}} \sum_{x_4 \in T \mathbb{Z} \backslash \{0\}}
\int \frac{d k_4}{2 \pi} \frac{1}{k^2} \frac{1}{(k+p)^2+m^2} e^{i k_4 x_4} \non
& \sim  e^{- |\vec{k}|_{\rm min} T} \sim  e^{- 2 \pi \frac{T}{L}} . \label{expsupp}
\end{align}
In the last line, we only showed the exponential factor of the contribution
which is given by the possible smallest  $|\vec{k}|$ and $|x_4|$.
($\int dk_4$ is performed by using the Cauchy theorem.)

\vspace{2mm}
\noindent
{\bf{QED$_{L}$ on $\mathbb{T}^4$
$\to$ 
QED$_{TL}$ on $\mathbb{T}^4$
}} \\
The difference between QED$_L$ on 
$\mathbb{T}^4$ and QED$_{TL}$ on $\mathbb{T}^4$ 
is given by
\begin{align}
\Delta'' I(L,T)
&=\lt(\SumInt_{\text{QED$_L$ on  $\mathbb{T}^4$}} - \SumInt_{\text{QED$_{TL}$ on  $\mathbb{T}^4$}}\rt) 
\frac{1}{k^2} \frac{1}{(k+p)^2+m^2} \non
&=-\frac{1}{L^3 T}\sum_{k_4 \in B \mathbb{Z}_T\backslash \{0\}} \frac{1}{k^2} \frac{1}{(k+p)^2+m^2} \bigg|_{\vec{k}=0} \non
&=-\frac{T}{L^3} \sum_{n \neq 0} \frac{1}{4 \pi^2 n^2} \frac{1}{\lt(\frac{2 \pi n}{T}+p_4 \rt)^2+m^2 } .
\end{align}
We may give a bound $|\Delta '' I| \leq \frac{T}{L^3} \frac{1}{m^2} \sum_{n \neq 0} \frac{1}{n^2}$.
The divergence of $\Delta'' I$ in the large-$T$ 
limit before the large-$L$ limit was known in ref.~\cite{Borsanyi:2014jba}.

\vspace{2mm}

The FV corrections are given by
$\Delta I+\Delta' I$ in QED$_L$ on 
$\mathbb{T}^4$,
and $\Delta I +\Delta' I+\Delta'' I$
in QED$_{TL}$ on $\mathbb{T}^4$.

We close this subsection with a remark regarding the relation
between $\Delta' I$ and $\Delta'' I$.
From eq.~\eqref{expsupp},  one can see that
$\Delta' I$ is exponentially suppressed for $2 \pi T/L \gg 1$
On the other hand, $\Delta'' I$ is eliminated
in the opposite hierarchy because of
$\Delta'' I \sim T/L^3$.
Therefore, if the large-$L$ limit is taken before
the large-$T$ limit, $\Delta' I$ is not suppressed exponentially
and the asymptotic behavior for $L \gg 1 $ needs to be clarified
(although the study of the asymptotic behavior is beyond the scope of this paper). In the next subsection, we focus on QED$_L$ and in particular
its FV effect, $\Delta I$, assuming
that the large-$T$ limit (with finite $L$) is taken, where $\Delta' I$
can be safely negelected.

\subsection{Finite volume corrections in QED$_L$: Euclidean
time correlator}
\label{sec:2.2}

We consider how large FV corrections in QED$_L$ appear in
the Euclidean time correlator, 
which is what one can obtain 
actually in lattice simulations:
\be
C(t)=\int \frac{d p_4}{2\pi}
\lt[ \frac{1}{p^2+m^2}
+\lt( \frac{1}{p^2+m^2} \rt)^2 m^2 e^2 (I_{\infty}(p^2)+\Delta I (p, L)) \rt] e^{i p_4 t} . \label{Ctwhole}
\ee
$I_{\infty}$ represents the infinite volume limit
result. $m^2$ in the numerator
of the second term is added
to compensate for the 
mass dimension.
This is a rough prescription,
but here we are only interested
in typical function form.

\begin{figure}[tb]
\begin{minipage}{0.32\hsize}
\begin{center}
\includegraphics[width=5.0cm]{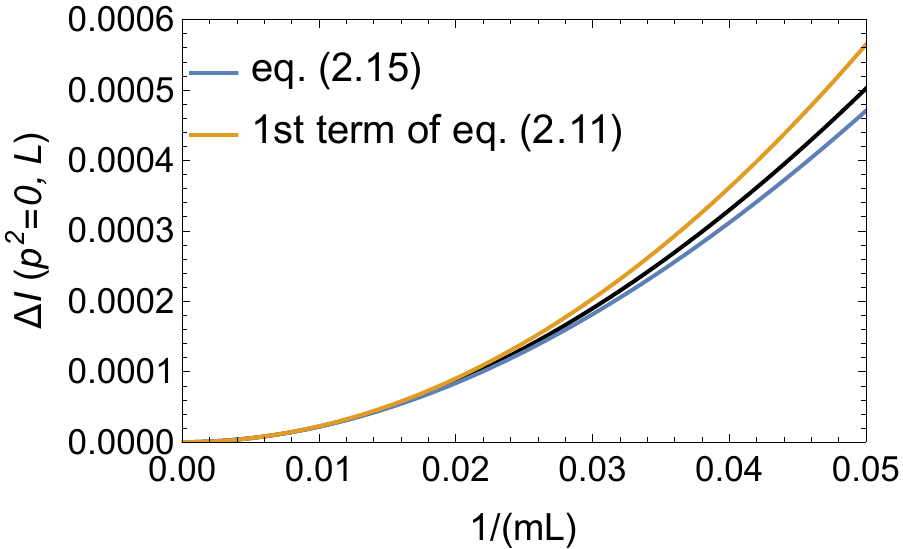}
\end{center}
\end{minipage}
\begin{minipage}{0.32\hsize}
\begin{center}
\includegraphics[width=5.0cm]{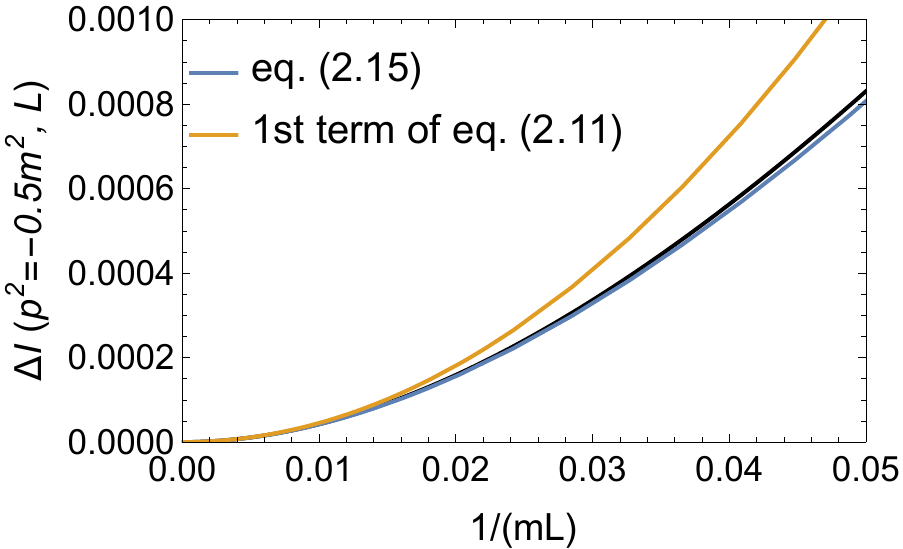}
\end{center}
\end{minipage}
\begin{minipage}{0.32\hsize}
\begin{center}
\includegraphics[width=5.0cm]{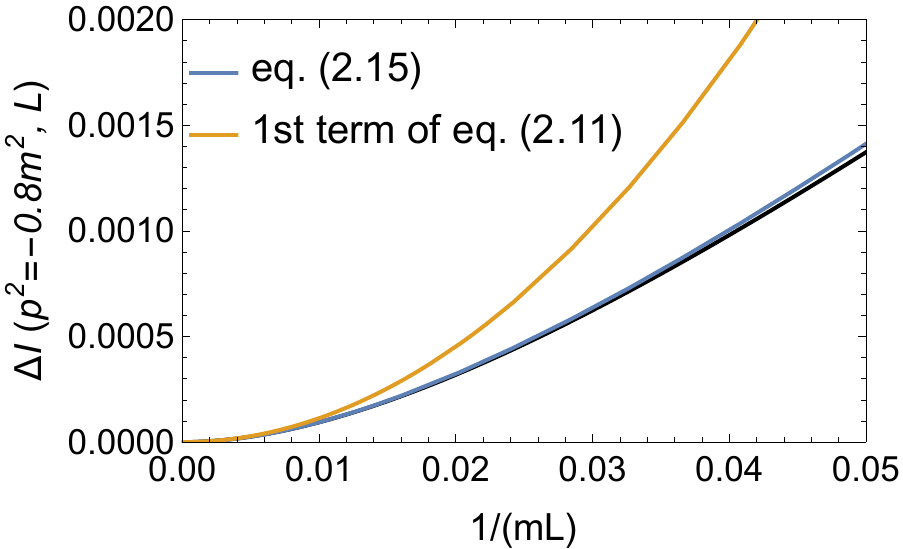}
\end{center}
\end{minipage}
\caption{Comparison of numerical evaluation of $\Delta I$ 
[eq.~\eqref{DeltaI}] (black)
with its asymptotic functions given by eq.~\eqref{modifiedDeltaI} (blue) and eq.~\eqref{EuclidDeltaI} (orange). They are shown  as a function of $1/(mL)$ and $p^2$ is chosen as $p^2=0$ (left), $p^2=-0.5 m^2$ (middle), and $p^2=-0.8 m^2$ (right).}
\label{fig:deltaI}
\end{figure}

Before we insert our result eq.~\eqref{EuclidDeltaI} into 
the above equation, 
we need to consider the function form of
$\Delta I$ in more detail.
It is known that
for the on-shell momentum $p_4=i m$
the FV correction is given by
$\Delta I (p=im,L) \sim 1/(mL)$ \cite{Borsanyi:2014jba, Davoudi:2018qpl}, which has a different $L$ dependence
from the off-shell case.
In our calculation, this can be understood 
by setting $p^2=-m^2$ in eqs.~\eqref{DeltaI} 
and \eqref{DeltaItilde};
then the $y$-integral in eq.~\eqref{DeltaItilde}
has the $u=-1$ singularity, which means
that $\Delta I \sim 1/(mL)$.
This difference can be understood 
from the fact that the IR behavior of 
the integrand of $I$ gets 
severer for the on-shell 
momentum than Euclidean momentum.
From these facts, we argue that
the FV correction is given by
\be
\Delta I(p, L)
=-\frac{\kappa_{-2}}{4 \pi} \frac{1}{p^2+m^2+c \frac{m^2}{m L}} \frac{1}{L^2} , \label{modifiedDeltaI}
\ee
with a constant $c$. 
This function form \eqref{modifiedDeltaI} is consistent 
with eq.~\eqref{EuclidDeltaI}
and also with the FV correction for the on-shell momentum. 
We note that $p^2=-m^2$ is not a singular point
and the pole position is slightly shifted to the point
$p^2=-(1+\frac{c}{mL}) m^2$ .
The constant is given by
$c=4 \kappa_{-3/2}/\kappa_{-2}=4$ such that
eq.~\eqref{modifiedDeltaI} correctly gives 
the FV correction for the on-shell momentum,
where\footnote{One can prove $\kappa_{-2}=\kappa_{-3/2}$
by rewriting the integrals $\int_0^{\infty} ds =[\int_0^1+\int_1^{\infty}] ds $  by $\int_0^1 ds$ integrals and using eq.~\eqref{thetarelation}.
}
\be
\kappa_{-3/2} \equiv \int_0^{\infty} ds \, s^{-3/2}  \lt(\vartheta_3 (0,e^{-\pi/s})^3-1-s^{3/2} \rt) =\kappa_{-2} .
\ee 
The validity of eq.~\eqref{modifiedDeltaI} in the complex $p^2$-plane is shown in fig.~\ref{fig:deltaI},
where we compare eq.~\eqref{modifiedDeltaI} with numerical evaluation of eq.~\eqref{DeltaI} for Euclidean momentum $p^2=0$ and momenta close to the mass shell. 

Inserting eq.~\eqref{modifiedDeltaI},
we obtain
\begin{align}
\int \frac{d p_4}{2 \pi}
\lt( \frac{1}{p^2+m^2} \rt)^2 m^2 \Delta I(p, L) e^{i p_4 t} 
&=\frac{e^{-mt}}{4 m}
\lt[(1+mt) \lt(-\frac{\kappa_{-2}}{4 \pi c} \rt) \frac{1}{mL} +\frac{\kappa_{-2}}{2 \pi c^2}\rt] \non
&\quad{}+\frac{e^{-m \sqrt{1+\frac{c}{ m L}}t}}{4 m}
\lt(-\frac{\kappa_{-2}}{2 \pi c^2} \frac{1}{\sqrt{1+\frac{c}{ m L}}} \rt) . \label{generalFV}
\end{align}
In the right hand side, the first line represents the contribution
from the double pole at $p_4=im$
and the second line the contribution from the pole
at $p_4=i m \sqrt{1+\frac{c}{m L}}$.

We see from eq.~\eqref{generalFV} 
that FV effect to
$C(t)$ or the effective mass
has different parametric 
dependences 
depending on the size of $t/L$.
(Here, $mL \gg 1$ is assumed and the time separation $t$ is implicitly taken to satisfy $mt \gg 1$ to discuss the long-distance regime.)
For $t/L \gg 1$, the second term
of eq.~\eqref{generalFV} is 
exponentially small $\sim e^{-c t/(2L)}$ 
compared with the first term
and can be neglected.
Then, in this case, the FV corrections to the effective mass
($M_{\rm eff}(t)=-\log[C(t)]'$) are given by 
\begin{align}
\Delta M_{\rm eff}(t)
=e^2 \frac{ \kappa_{-2}}{8 c \pi } \frac{1}{L}+\cdots.
\end{align}
The above FV correction of $\sim 1/L$ can be correctly 
read off from $\Delta I (p_4= im,L) \sim 1/(mL)$, 
where the on-shell momentum is assumed 
from the beginning in calculating
the momentum-space correlator.

On the other hand, for $t/L \ll 1$, one can expand 
the exponential factor in the second term
of eq.~\eqref{generalFV} in $t/L$,
which leads to
\be
({\rm Eq.}~\eqref{generalFV})
=-\frac{e^{-mt}}{2 m}
\frac{\kappa_{-2}}{32 \pi} 
(3+3 m t +m^2 t^2) \frac{1}{(mL)^2}+\cdots .
\ee
The $1/(mL)$ terms 
are {\it{canceled}} between 
the first and the second contributions in eq.~\eqref{generalFV}. 
In this case the FV corrections to the effective mass
are given by
\be
\Delta M_{\rm eff}(t)
=e^2 \frac{ \kappa_{-2}}{32 \pi } \frac{2 m t+3}{m L^2}+\cdots . \label{Meffsmallt}
\ee
The above result can be also
obtained if we simply use eq.~\eqref{EuclidDeltaI}.
(In this case, note that $p^2=-m^2$ is a triple pole in the integand of eq.~\eqref{Ctwhole}.)
We note that this result cannot be obtained
if one only knows $\Delta I (p=im,L) \sim 1/(mL)$
and that the off-shell result of $\Delta I$ is essential.

For $t/L \ll 1$, 
the FV effect is parametrically
smaller than the case of 
$t/L \gg 1$
but gives a time-dependent effective mass as above.
For $t/L \gg 1$, the FV effect is larger but it is not time dependent.

The fact that the FV effect is smaller for $t/L \ll 1$
may motivate one to focus on the $1/m \ll t \ll L$ region.
In this case one should note that
the effective mass is affected by
the contribution from the branch cut starting from
$p^2=-m^2$, corresponding to
the particle production cut.
This drawback could be overcome
by removing in advance 
the branch cut effect,
which is calculable, from 
the effective mass.
We calculate the branch
cut effect in App.~\ref{sec:A}
in the scalar QED in infinite volume, 
and show that 
the cut effect can be largely removed from the effective mass.

We make comments on the connection with 
ref.~\cite{Davoudi:2018qpl}, where
FV corrections in scalar QED 
have been studied in detail, and the FV effects to the pole mass of $O(1/(mL))$ are confirmed. 
In extracting the pole mass from their lattice data, a clear plateau in the effective mass has not been observed.
They then have performed a subtraction of the contributions from the multi-particle states (collection of poles with $1/L$ intervals), after which a good plateau has been successfully obtained.
They have read off the pole mass from the plateau and repeat the calculation for various $L$, which agrees with their expectation of the $O(1/(mL))$ behavior.

This conclusion is perfectly consistent with our discussion but with a little bit of complication.
As discussed above, the effective mass $M_{\rm eff} (t)$ is schematically expressed as the sum of the contributions from the pole at $p^2 = -m^2$ (labeled as `pole') and those from the multi-particle states (labeled as `cut') each with FV effects labeled as $\Delta$ as follows:
\begin{align}
    M_{\rm eff} (t) =
    M_{{\rm eff}, \infty}^{\rm pole}
    + M_{{\rm eff}, \Delta}^{\rm pole}
    + M_{{\rm eff}, \infty}^{\rm cut} (t)
    + M_{{\rm eff}, \Delta}^{\rm cut} (t)
    + \cdots,
\end{align}
where $\cdots$ represents the higher order terms in the QED perturbations. The non-trivial $t$ dependence appears from the cut contributions such as $M_{{\rm eff}, \infty}^{\rm cut} (t) \sim \alpha /t$. (This behavior can be understood from the log term in eq.~\eqref{Ctcuttilde}.)
The FV effects in the cut contributions have been already discussed above and the large $L$ behavior are found that 
\be
    M_{{\rm eff}, \Delta}^{\rm cut} (t) \sim \bigg\{
    \begin{array}{ll}
    {c_1 \alpha \over L} + {c_2 \alpha t \over L^2}, & \quad t \ll L,\\
    \alpha c_3 m e^{-c_4 t/L}, & \quad t \gg L,
   \end{array}
   \label{eq:meffdeltacut}
\ee
with coefficients of $O(1)$, $c_{1-4}$. We also find that the FV contribution from the pole in the large volume limit is given by
\begin{align}
    M_{{\rm eff}, \Delta}^{\rm pole} \sim - {c_1 
    \alpha \over L},
\end{align}
where $c_1$ is common to that in Eq.~\eqref{eq:meffdeltacut}. This is understood from Eq.~\eqref{Meffsmallt} where there are no $O(1/L)$ terms for $t \ll L$.
In ref.~\cite{Davoudi:2018qpl}, the subtraction is done as 
$M_{{\rm eff}} (t) 
- (M_{{\rm eff}, \infty}^{\rm cut} (t)
 + M_{{\rm eff}, \Delta}^{\rm cut} (t))$,
 which results in the FV effects of $O(1/L)$ from $M_{{\rm eff}, \Delta}^{\rm pole}$.
 
 However, if we subtract only $M_{{\rm eff}, \infty}^{\rm cut} (t)$, which is analytically calculable, the FV effects are obtained to be $O(t/L^2)$ by the cancellation of the $O(1/L)$ term in the combination of $M_{{\rm eff}, \Delta}^{\rm pole} + M_{{\rm eff}, \Delta}^{\rm cut} (t)$ for $t \ll L$. This FV effect is parametrically smaller than $O(1/L)$ and, in addition, the non-plateau behavior could be largely improved by subtracting $M_{{\rm eff}, \infty}^{\rm cut}$ alone, as indicated in App.~\ref{sec:A}.
 If one applies this strategy in actual lattice computations, the FV effects may be further controlled.

We note that our study is performed with a simple and rough integrand $\frac{1}{k^2[(k+p)^2+m^2]}$ whereas in ref.~\cite{Davoudi:2018qpl} the realistic integrand appearing in scalar QED is considered.
We also note, however, that the same denominator is considered,
which essentially determines parametric dependence of FV corrections. Therefore, the simple integrand assumed in our study is sufficient to highlight 
the important point of our discussion.
Nevertheless, it would be useful 
to study the FV corrections to the effective mass by starting with the realistic integrand and to
give a result like eq.~\eqref{Meffsmallt}.
Then it could be possible to
control FV corrections more easily.
In this attempt, the technique developed in ref.~\cite{Davoudi:2018qpl} might be useful.

We have observed that the FV effect in QED$_L$ is sensitive to the IR structure (as discussed above eq.~\eqref{modifiedDeltaI})
 and exhibits a complicated $t$ dependence in the time correlator.
In our computation aiming at obtaining the $g$ factor {\it{to high orders}},
it is not realistic to correctly understand IR structure of the relevant three-point function at each loop-expansion order
and determine a proper fit function to remove FV corrections.
Consequently, we find QED$_L$ unsuited for our purpose,
and, therefore, consider another regularization method.

\subsection{Massive photon regularization}
\label{sec:2.3}
We consider massive photon regularization with photon mass of $m_{\gamma} L \gg 1$.
In this case, the FV effects are known to be exponentially suppressed.
Let us see this explicitly for clarity.
$I$ is evaluated as
\be
I=\frac{1}{16 \pi^2}\sum_{x \in L \mathbb{Z}^3 \times T \mathbb{Z}}  \int_0^1 dy \, \int_0^{\infty} ds \, s^{-1} 
\exp\lt[-   \lt\{s [y(1-y)p^2+y m^2+(1-y) m_{\gamma}^2]+\frac{|x|^2}{4 s}  \rt\}-i y p \cdot x  \rt] .
\ee
The FV effect, $\Delta I$, is given by the contributions where $x \neq 0$.
We obtain a bound
\begin{align}
&\lt| \int_0^{\infty} ds \, s^{-1} \exp\lt[-   \lt\{s [y(1-y)p^2+y m^2+(1-y) m_{\gamma}^2]+\frac{|x|^2}{4 s}  \rt\}-i y p \cdot x  \rt] \rt| \non
&\leq  \int_0^{\infty} ds \, s^{-1} \exp\lt[-   \lt\{s  m_{\gamma}^2+\frac{|x|^2}{4 s}  \rt\} \rt]
= 2 K_0( m_{\gamma} |x|) 
<\frac{C}{\sqrt{m_{\gamma} |x|}} \exp[-m_{\gamma} |x|] .
\end{align}
$K_0$ is the Bessel function of second kind.
There exists a constant $C$ satisfying the above inequality.
In the above calculation, we assumed 
$m^2 > m_{\gamma}^2$, so that
$y m^2+(1-y) m_{\gamma}^2 \geq m_{\gamma}^2$.
Using the above bound, we can see that
the FV effects are suppressed as
$\sim e^{-m_{\gamma} L}, e^{-m_{\gamma} T}$ 
by focusing on the smallest $|x|$
contributions, which dominate $\Delta I$.

Therefore, as long as we take $m_{\gamma} L, m_{\gamma} T \gg 1$,
we can safely neglect FV effects on Euclidean quantities.
As can be seen from the above calculations, this is true independent of 
choices of the external momentum.

Since we can expect
exponential suppression of 
FV corrections at any loop orders and 
any correlation functions, we will adopt finite photon mass
regularization in our computation.

\section{Lattice calculation of $g-2$}
\label{sec:3}

In this section, we 
perform perturbative computation of 
the electron $g$ factor on the lattice.
We adopt photon
mass regularization.
In Sec.~\ref{sec:3.1}, we explain the outline of our calculation of
 the $g$ factor.
 In Sec.~\ref{sec:3.2}, we study
 backward propagation and 
 propose a method to suppress its effects.
In Sec.~\ref{sec:3.3}, we study corrections to the $g$ factor caused by finite photon mass, finite photon momentum, and finite smearing parameter,
which are introduced
in our calculation (but 
should be finally sent to zero
or infinity).
In Sec.~\ref{sec:3.4}, we
then discuss an optimal strategy for extracting
the $g$ factor based on 
our understanding of 
systematic errors.
In Sec.~\ref{sec:3.5}, 
we carry out numerical simulation
following the strategy presented in
Sec.~\ref{sec:3.4}.

\subsection{Outline of the method}
\label{sec:3.1}

Our calculation is based on ref.~\cite{Kitano:2021ecc} to a large extent,
but in order to clarify some modifications and to define quantities necessary for the discussion below,
we briefly explain our calculation method. 
We consider the quenched QED action on the Euclidean lattice, setting the lattice spacing $a$ to $a=1$, as
\begin{align}
S
&=\frac{1}{4}  \sum_{n, \mu, \nu} \lt[e^{-\nabla^2/\Lambda_{\rm UV}^2} (\nabla_{\mu} A_{\nu}(n)-\nabla_{\nu} A_{\mu}(n) ) \rt]^2 \non
&\quad{}+\frac{1}{2 \xi} \sum_{n} \lt[e^{-\nabla^2/\Lambda_{\rm UV}^2} \sum_{\mu} \nabla^{*}_{\mu} A_{\mu}(n) \rt]^2
+\frac{1}{2} m_{\gamma}^2 \sum_{n,\mu} [e^{-\nabla^2/\Lambda_{\rm UV}^2} A_{\mu}(n)]^2 ,
\label{eq:action}
\end{align}
where 
\be
\nabla_{\mu} f(n) \equiv f(n+\hat{\mu})-f(n), \quad{} \nabla^*_{\mu} f(n) \equiv f(n)-f(n-\hat{\mu}) ,
\ee
and $\nabla^2=\sum_{\mu} \nabla_{\mu} \nabla^*_{\mu}$.
Here we introduce a smearing parameter $\Lambda_{\rm UV}$,
finite photon mass $m_{\gamma}$, and a gauge fixing parameter $\xi$ \cite{Kitano:2021ecc}.
The photon two point function is exactly given in quenched QED as
\begin{align}
\langle \tilde{A}_{\mu}(k) \tilde{A}_{\nu}(k') \rangle
&=V \delta_{k+k',0} e^{-2 \hat{k}^2/\Lambda_{\rm UV}^2} \lt[\lt(\delta_{\mu \nu}-\frac{\hat{k}_{\mu} \hat{k}_{\nu}}{\hat{k}^2} \rt) 
\frac{1}{\hat{k}^2+m_{\gamma}^2}+\frac{\hat{k}_{\mu} \hat{k}_{\nu}}{\hat{k}^2} \frac{\xi}{\hat{k}^2+\xi m_{\gamma}^2} \rt] \non
&\equiv V \delta_{k+k',0}  \mathcal{D}_{\mu \nu}(k)
\label{eq:photon_propagator}
\end{align}
where $\hat{k}_{\mu} \equiv 2 \sin(k_{\mu}/2)$ and $\hat{k}^2=\hat{k}_{\mu} \hat{k}_{\mu}$.
$V$ denotes the four-dimensional spacetime volume.
We define the Fourier transform by $\tilde{A}_{\mu}(k)=\sum_{n} A_{\mu}(n) e^{-ik (x_n+\hat{\mu}/2)}$.
There are no quantum corrections to $\xi$, $m_{\gamma}$, and 
the normalization of the photon wave function in quenched QED. 
Even though we introduce non-zero photon mass, we argued \cite{Kitano:2021ecc} that 
similar Ward-Takahashi (WT) identities to the massless photon theory hold.
We can repeat a parallel argument in the quenched QED case.
Particularly we can conclude that the renormalization of the coupling constant is related to the wave function renormalization of the photon
as
\be
e_P=Z_3^{1/2} e .
\ee
As noted, $Z_3$ can be exactly calculated at the tree-level due to the absence of the dynamical fermion.

We are interested in the $g$-factor, which can be extracted from the three-point function
\begin{align}
G_{\mu}(p,k) 
&\equiv \frac{1}{V}  \sum_{n,m,\ell} \langle (D^{-1})_{nm} A_{\mu}(\ell) \rangle e^{-i p x_n} e^{-i (-p-k) x_m} e^{-ik(x_{\ell}+\mu/2)} \non
&=\frac{1}{V} \langle  \tilde{D}^{-1}(p,-p-k) \tilde{A}_{\mu}(k) \rangle ,
\label{eq:3point}
\end{align}
where the photon momentum is $k$, and the ingoing and outgoing fermion momenta are $p$ and $p+k$, respectively.
Here the matrix $D$ is the covariant derivative acting to the fermion field in the nonquenched QED Lagrangian and given by
\be
D_{n m} \equiv m \delta_{nm}+\frac{1}{2} \sum_{\mu} \left[\gamma_{\mu} e^{ie A_{\mu}(n)} \delta_{n+\hat{\mu},m}
-\gamma_{\mu} e^{-ie A_{\mu}(n-\hat{\mu})} \delta_{n-\hat{\mu},m}\right]  \label{covderi}.
\ee
We denote the bare electron mass by $m$
and the on-shell mass by $m_f$.
We define the electron propagator by
\be
S(p) \equiv \frac{1}{V} \langle \tilde{D}^{-1}(p,-p)  \rangle .
\ee
We define the vertex function $\Gamma_{\mu}(p,k)$ amputating the external legs,
\be
- i e_P \Gamma_{\mu}(p,k) =\kappa \mathcal{D}^{-1}_{\mu \nu}(k) S(p)^{-1} G_{\nu}(p,k) S(p+k)^{-1} ,
\ee
where $\kappa$ is given by a product of renormalization factors but needs not to be specified here.
In continuum spacetime, $\bar{u}(p) \Gamma_{\mu}(p,k) u(p+k)$ can be decomposed into two parts:
\be
- i e_P \bar{u}(p) \Gamma_{\mu}(p,k) u(p+k) 
=- i e_P  \bar{u}(p) \lt( F_1(k^2) \gamma_{\mu}-F_2(k^2) \frac{\sigma_{\mu \nu} k_{\nu}}{2 m_f} \rt) u(p+k)
\ee
where $\sigma_{\mu \nu}=(i/2) [\gamma_{\mu}, \gamma_{\nu}]$ and the wave function satisfies $(-i \Slash{p}-m) u(p)=0$
with $p_0=i\sqrt{\vec{p}^2+m_f^2}$.
The $g$ factor is defined by
\be
\frac{g}{2}=\frac{F_1(0)+F_2(0)}{F_1(0)} .
\ee

To obtain the $g$ factor on the lattice, we need to compute 
\be
\hat{G}_{\mu}(p,k) \equiv \mathcal{D}^{-1}_{\mu \nu}(k) G_{\nu}(k,p) \label{Ghat},  
\ee
and 
\be
\hat{G}^{(\text{norm})}_{\mu}(p,k)=-i S(p) \gamma_{\mu} S(p+k) \label{Gnorm} .
\ee
Although the form factors are defined for the {\it on-shell} fermion,
we obtain these quantities for Euclidean momenta.
To read off on-shell amplitudes, we consider Fourier transform of $\hat{G}_{\mu}(p,k)$ and $\hat{G}^{(\text{norm})}_{\mu}(p,k)$.
The formula we use is
\be
\frac{g}{2}=\lim_{t \to \infty} \frac{g(t)}{2}, \quad {g(t) \over 2}=\frac{\mathcal{F}_M(t)/\mathcal{F}_E(t)}{\mathcal{F}_M^{(\text{norm})}(t)/\mathcal{F}_E^{(\text{norm})}(t)}, \label{gfacmethod}
\ee
where 
\be
\mathcal{F}_E(t)=\sum_{p_4} {\rm tr} [\gamma_4 \hat{G}_4] e^{i p_4 t} , \label{FE}
\ee
\be
\mathcal{F}_M(t)=\sum_{p_4} \sum_{i,j,k=1}^3 i \epsilon_{ijk} {\rm tr} [\gamma_5 \gamma_i \hat{G}_j] \hat{k}_k e^{i p_4 t} , \label{FM}
\ee
and $\mathcal{F}^{(\text{norm})}_E(t)$ and $\mathcal{F}^{(\text{norm})}_M(t)$ are defined in a parallel manner by using
$\hat{G}^{(\text{norm})}$ instead of $\hat{G}$.
The reason why we can obtain the $g$ factor using this formula is explained in Sec.~6 in ref.~\cite{Kitano:2021ecc}.
We note that systematic uncertainties,
such as discretization effects and finite
photon momentum effects, are left in
eq.~\eqref{gfacmethod} and should be
removed.

We explain how to compute $\hat{G}_{\mu}(p,k)$. From the fact that
\be
\int D A_{\mu} \, (D[A_{\mu}]^{-1})_{nm} e^{-S[A_{\mu}]}
\ee
is invariant under the change of the integration variable $A_{\mu}(n) \to A_{\mu}(n)+\epsilon_{\mu}(n)$ 
(here we explicitly show that the covariant derivative \eqref{covderi} is a functional of $A_{\mu}$),
we obtain a Schwinger-Dyson equation,
\be
-\lt\langle  \frac{\delta S}{\delta A_{\mu}(\ell)} D^{-1}_{nm} \rt\rangle
=\lt\langle \lt( D^{-1} \cdot \frac{\delta D}{\delta A_{\mu}(\ell)} \cdot D^{-1} \rt)_{nm} \rt\rangle .
\ee
Noting that the left-hand side is given by
\be
-\sum_{k'} \frac{1}{V} e^{i k' (x_{\ell}+\hat{\mu}/2)} \mathcal{D}^{-1}_{\mu \nu} (k') \lt\langle \tilde{A}_{\nu}(k') (D^{-1})_{nm}  \rt\rangle ,
\ee
we obtain an identity in momentum space,
\be
\hat{G}_{\mu}(p,k)=-\frac{1}{V} \sum_{n,m, \ell}  \lt\langle \lt( D^{-1}
\cdot \frac{\delta D}{\delta A_{\mu}(\ell)} \cdot D^{-1} \rt)_{nm}
\rt\rangle e^{-i p x_n} e^{-i (-p-k) x_m}
e^{-ik(x_{\ell}+\hat{\mu}/2)} \label{SDeq} .
\ee
In giving $\hat{G}_{\mu}(p,k)$, we evaluate
the right-hand side, replacing
$\delta D/\delta A_{\mu}(\ell)$ with its
$a \to 0$ limit value. The use of
eq.~\eqref{SDeq} significantly
reduces the statistical error. 
The computation via
Eq.~\eqref{eq:3point} at the fixed order reduces to calculating
multi-point correlation functions of photons, where the momentum
conservation requires to pick up $\tilde A_\mu (k)$ in the expansion
of $\tilde D^{-1}$. This is done only statistically while the above
formula can skip the procedure.

We evaluate eqs.~\eqref{Ghat} [or \eqref{SDeq}] and \eqref{Gnorm} in perturbation theory.
The perturbative formulae for $\delta D/\delta A_{\mu}$ and $D^{-1}$ are given in Sec.~2 of ref.~\cite{Kitano:2021ecc}.
We note that, as explained in ref.~\cite{Kitano:2021ecc}, $D^{-1}$ can be evaluated fast 
within perturbation theory by using the fast Fourier transform. 
After the formal perturbative expansion of $D^{-1}$ and $\delta D/\delta A_{\mu}$,
all we have to do is to evaluate correlation functions of gauge fields,
$\langle A_{\mu_1}(n_1) A_{\mu_2}(n_2) \dots \rangle$. Since the weight $e^{-S[A_{\mu}]}$
is just a Gaussian, we generate configurations according to the Gaussian distribution
to measure expectation values.

In our lattice calculations, we consider momenta,
\be
p=(\vec{p},p_4)=(-\vec{k}/2, p_4 )
\ee
with $k=(0,0,2 \pi/L,0)$ and  $(2 \pi/L ,0,2\pi/L,0)$. 
These two momenta are used for extrapolation
to $k^2 \to 0$.
 $p_4$ is summed over afterwards in eq.~\eqref{FE} etc.
For this choice, the other fermion has momenta
\be
p+k=(\vec{k}/2, p_4)
\ee
and the positions of the on-shell pole of $S(p)$ and $S(p+k)$ are the same.

\subsection{Backward propagation and its suppression}
\label{sec:3.2}

The sum over $p_4$ such as the one in eq.~\eqref{FE}
gives a contribution from the propagation which wraps around
the torus (backward propagation).
For example, let us consider
\be
f_T(t) \equiv \frac{1}{T} \sum_{p_4} \frac{1}{\sin \lt( p_4 \rt)^2+E^2} e^{i p_4 t}
=\frac{1}{T} \sum_{n=0}^{T-1} \frac{1}{\sin \lt(\frac{2 \pi}{T} n \rt)^2+E^2} e^{i \frac{2 \pi}{T} n t} ,
\ee
which mimics Fourier transform of a two point function, where $E$ denotes energy.
We obtain the exact formula for this sum,
\be
f_T(t)
=\frac{4}{\lt(\frac{1}{z_*^2}-z_*^2 \rt) (1-z_*^T) } (z_*^t+z_*^{T-t})  \label{backward}
\ee
for even $t$ with $0 \leq t < T$ and zero for odd $t$.
$z_*$ denotes a pole in the $z(=e^{i p_4})$-plane and is given by $z_*=- E+\sqrt{1+E^2}$.

In eq.~\eqref{backward}, $z_*^{T-t}$ is the backward propagation.
This is an obstacle in investigating the large-$t$ behavior
because it becomes similar in size to the interested contribution $z_*^t$
for large $t$. 
This was actually an obstacle in our previous study~\cite{Kitano:2021ecc}.

Let us consider to take finer $p_4$ in the sum,
\be
g(t) \equiv \frac{1}{2 T} \sum_{n} \frac{1}{\sin \lt(\frac{2 \pi}{T} n \rt)^2+E^2} e^{i \frac{2 \pi}{T} n t}
\ee
with $n=0, 1/2, 1, 3/2, \dots$. One can see that $g(t)=f_{2T}(t)$.
Therefore, in this case the backward propagation is given by 
$z_*^{2T -t}$ and gets milder than the previous case. 

We suppress the backward propagation in our study by taking finer $p_4$ as in the above case.
We accomplish this by changing boundary conditions for the
fermion~\cite{Bedaque:2004kc, DelDebbio:2018ftu}, i.e., we 
employ the periodic and anti-periodic boundary conditions.

\subsection{Effects of $m_{\gamma}$, $k^2$, and $\Lambda_{\rm UV}^2$}
\label{sec:3.3}

The $g$ factor obtained by
eq.~\eqref{gfacmethod} receives
discretization effects, 
finite volume effects,
and effects of modification 
of the Lagrangian, i.e., 
finite photon mass and finite
$\Lambda_{\rm UV}^2$,
and also non-zero photon momentum $k^2$. 
(Although $g$ factor is defined 
for $k^2 \to 0$, we consider finite 
$k^2$ in our calculation.)
Here we consider effects of
$m_{\gamma}$, 
$k^2$ and $\Lambda_{\rm UV}^2$ 
in {{\it{continuum and infinite volume }} spacetime
because this is sufficient to evaluate their dominant effects.
The FV corrections are 
exponentially suppressed for 
$m_{\gamma} L \gg 1$
as discussed in 
Sec.~\ref{sec:2.3}.
The discretization effects 
are discussed in the 
subsequent subsection. 

Our consideration is given at the one-loop level. At one loop, it is sufficient to focus only on $F_2$ because the one-loop contribution
to $F_1$ totally vanishes in the quantity $(F_1(k^2)+F_2(k^2))/F_1(k^2)$ due to 
$F_1=1+\mathcal{O}(\alpha)$ and $F_2=\mathcal{O}(\alpha)$.
Assuming the on-shell fermion momenta $p^2=(p+k)^2=-m_f^2$, we have
\begin{align}
&\bar{u}(p) \Gamma(p,k ) u(p+k) |_{\text{one-loop}} \non
&=-e^2 \int \frac{d^d \ell}{(2\pi)^d} \bar{u}(p) \frac{\gamma_{\rho} (-i (\Slash{p}+\Slash{\ell}) +m_f)\gamma^{\mu} (-i(\Slash{p}+\Slash{k}+\Slash{\ell})+m_f) \gamma^{\rho}}{[m_f^2+(p+\ell)^2][m_f^2+(p+k+\ell)^2] [\ell^2+m_{\gamma}^2]} u(p+k) \non
&=-e^2 \int_0^1 dy dz 2 y\int \frac{d^4 \ell}{(2\pi)^4} \bar{u}(p) \frac{4 \Slash{\ell} \bar{p}^{\mu}-4i m_f \ell^{\mu}-2(2-d) \Slash{\ell} \ell^{\mu}}{[\ell^2+y \ell \cdot (\bar{p}+(1-2z)k)+(1-y) m_{\gamma}^2]^3} u(p+k)+(\gamma_{\mu}\text{-term}) .
\end{align}
We expressed $p$ and $k$ by linear combinations of $\bar{p} \equiv (p+k)+p$ and $k=(p+k)-p$.
They satisfy $\bar{p} \cdot k=0$ and $\bar{p}^2=-4 m_f^2-k^2$ for $p^2=(p+k)^2=-m_f^2$.
After some calculation and using the Gordon identity $\bar{u}(p)(-i \bar{p}_{\mu}) u(p+k)=\bar{u}(p) (2 m_f \gamma_{\mu}+ \sigma_{\mu \nu} k_{\nu} )u(p+k)$, we obtain
\begin{align}
&\bar{u}(p) \Gamma(p,k ) u(p+k) |_{\text{one-loop}} \non
&=- e^2 \bar{u}(p) \frac{\sigma_{\mu \nu} k_{\nu}}{2 m_f} u(p+k) \frac{1}{(4 \pi)^2} \int dy dz \frac{4 y^2(1-y)}{y^2+y^2 z(1-z) \frac{k^2}{m_f^2}+(1-y) \frac{m_{\gamma}^2}{m_f^2}}
+(\gamma^{\mu}\text{-term})  \non
&=- e^2 \bar{u}(p) \frac{\sigma_{\mu \nu} k_{\nu}}{2 m_f} u(p+k)  \frac{1}{(4\pi)^2} \lt(2-\frac{1}{3}\frac{k^2}{m_f^2}-2\pi \frac{m_{\gamma}}{m_f}+\cdots \rt)
+(\gamma^{\mu}\text{-term})  .
\label{finitemassk}
\end{align}
In the final expression, we assumed $k^2/m_f^2 \ll1 $ and $m_{\gamma}^2/m_f^2 \ll 1$.
Inside the brackets, the first term ``$2$'' gives the exact result of $F_2(0)$.
The finite $k^2$ and $m_{\gamma}^2$ effects are found to be $\mathcal{O}(k^2/m_f^2)$ and $\mathcal{O}(m_{\gamma}/m_f)$.

When we also turn on the UV cutoff scale $\Lambda_{\rm UV}$, 
we have
\begin{align}
&\bar{u}(p) \Gamma(p,k ) u(p+k) |_{\text{one-loop}} \non
&=-e^2 \frac{1}{16 \pi^2} \bar{u}(p) \frac{\sigma_{\mu \nu} k_{\nu}}{2 m_f} u(p+k) \non
& \times \int_0^1 dy d z \, 
\bigg[ 4 y^2   \int_0^{\infty} ds \,  \frac{s^3}{\lt(s+\frac{2 m_f^2}{\Lambda_{\rm UV}^2} \rt)^3} e^{-\frac{s^2}{s+\frac{2 m_f^2}{\Lambda_{\rm UV}^2}} y^2 (1+z(1-z) k^2/m_f^2)-s (1-y) m_{\gamma}^2/m_f^2}  \non 
&\qquad{}\qquad{}\quad{}-4 y^3 \int_0^{\infty} ds \,  \frac{s^4}{\lt(s+\frac{2 m_f^2}{\Lambda_{\rm UV}^2} \rt)^4} e^{-\frac{s^2}{s+\frac{2 m_f^2}{\Lambda_{\rm UV}^2}} y^2 (1+z(1-z) k^2/m_f^2)-s (1-y) m_{\gamma}^2/m_f^2}\bigg]  \non
&\quad{} +(\gamma^{\mu}\text{-term}) .
\end{align}
This additionally gives a correction of $\mathcal{O}[(2 m_f^2/\Lambda_{\rm UV}^2) \log{(2 m_f^2/\Lambda_{\rm UV}^2)}]$.\footnote{
The $s$-integral can be decomposed into $\int_0^{\infty}ds=\int_0^1 ds+\int_1^{\infty} ds$.
While the integral $\int_1^{\infty} ds$ gives
an expansion in $2 m_f^2/\Lambda_{\rm UV}^2$,
the integral $\int_0^1 ds$ can give a
$2 m_f^2/\Lambda_{\rm UV}^2 \log(2 m_f^2/\Lambda_{\rm UV}^2)$ term.}

\subsection{Calculation strategy}
\label{sec:3.4}

In this section, we discuss an optimal strategy to obtain the $g$
factor based on the above studies. As an IR regularization, as
mentioned above, we adopt photon mass regularization. We consider the
following properties advantageous. (i) It is clear that the finite
volume effects are exponentially suppressed, independent of the
details of considered quantities. 
(ii) Owing to (i),
the IR structure of massive photon theory with finite lattice is the
same as massive photon theory with infinite volume. The latter is well
understood.

Adopting massive photon regularization, we
first consider $k^2 \to 0$ extrapolation.
In the $g$ factor, IR divergences appear only for $k^2 \neq 0$ (when $m_{\gamma} =0$). 
This extrapolation therefore removes the IR divergences and makes the $g$ factor nonsingular at $m_{\gamma}=0$.
Since we found the finite $k^2$ effect to be $\mathcal{O}(k^2/m_f^2)$ in the one-loop analysis,
we perform linear extrapolation in $k^2/m_f^2$.

At this stage, we are left with finite photon mass effects and
finite lattice spacing effects.
We assume finite photon mass effects to be $\mathcal{O}(m_{\gamma}/m_f)$
from the above study.
Finite lattice spacing effects are given by $\mathcal{O}(m_f^2 a^2)$.
We also expect finite lattice spacing effects of $\mathcal{O}(m_{\gamma} a )$
because we do not find a reason to prohibit this for finite $m_{\gamma}$.
In this sense, it is straightforward to perform $m_{\gamma} \to 0$ extrapolation
before $m_{f} \to 0$ extrapolation.
To summarize, we first perform
linear extrapolation in $k^2/m_f^2 \to 0$,
secondly linear extrapolation in $m_{\gamma}/m_f \to 0$,
and finally linear extrapolation
in $m_f^2 a^2 \to 0$.

Our concrete setup is as follows.
For one lattice size $L^3 \times T$, 
we choose one fermion mass $m$.
We use two different
photon momenta to perform $k^2 \to 0$ extrapolation,
and also vary $m_{\gamma}$ to perform
$m_{\gamma}/m_f \to 0$ (and $m_{\gamma} a \to 0$)
extrapolations. 
For a different lattice size, we choose a different
fermion mass. 
Different lattices are used to perform
$m_f^2 a^2 \to 0$ extrapolation.
 (See Table~\ref{tab:lattice}.)

If we aims at 10~\% precision lattice data (before any extrapolations)
while setting
\be
m_{\gamma} L \simeq 4
\ee
in order to sufficiently suppress FV effects,
we need\fn{
We neglect $m_{\gamma} a$ in this discussion.}
\be
k_{\rm min}^2/m_f^2, \, m_{\gamma}/m_f, \, m_f^2 a^2 \lesssim 0.1 .
\ee
These conditions are satisfied with $L \gtrsim 120$.
This large (but realistic) lattice is required for precision study.

We finally mention finite $\Lambda_{\rm UV}^2$
effects.
In our simulation, we use a fixed 
value of $\Lambda_{\rm UV}^2 a^2$; 
see Table~\ref{tab:lattice}.
Since we use larger fermion mass for larger lattice,
finite $\Lambda_{\rm UV}^2$ effects, $\sim 2 m_f^2/\Lambda_{\rm UV}^2$, 
are expected to be removed simultaneously 
in the $m_f^2 a^2 \to 0$ extrapolation.

\subsection{Numerical results}
\label{sec:3.5}

We carry out lattice simulation 
following the strategy discussed
above.
However, unfortunately since we 
cannot use large  enough lattice
in the present study, 
we cannot keep all the systematic
uncertainties under good control.
We nevertheless perform 
lattice simulation to test feasibility of the calculation.

We perform a
numerical simulation by using small lattices listed in
Table~\ref{tab:lattice}.
For each lattice size, the smallest $m_\gamma$ is chosen so that
$m_\gamma L = 4.0$ which makes FV effects exponentially
suppressed. The fermion mass $m$ and other values of $m_\gamma$ are
then chosen to satisfy $m_\gamma /m = 0.4, 0.5, 0.6, 0.7$. These
points are used for taking the $m_\gamma \to 0$ limit.
(These points are not small enough to reliably perform the extrapolation to $m_{\gamma} \to 0$.
This limitation comes from the size of $L$, which we want to take bigger in our future work.)

\renewcommand{\arraystretch}{1.06}
\begin{table}[tb]
  \begin{center}
    \begin{tabular}[t]{c|c|c|c|c|r|r}
       $L^3 \times T$ & $ma$ & $(\Lambda_{\rm UV}a)^2 $ & $\xi$ & $m_\gamma a$ &
       $N_{\rm conf}$ & extrapolation points\\ \hline
       $14^3 \times 28$ & $0.714$ & $4.0$ & $1.0$ & $0.2857, 0.3571,
       0.4286, 0.50$ & 780 & [7:9]\\
       $16^3 \times 32$ & $0.625$ & $4.0$ & $1.0$ & $0.25, 0.3125,
       0.375, 0.4375$ & 780 & [7:9]\\
       $18^3 \times 36$ & $0.556$ & $4.0$ & $1.0$ & $0.222, 0.278, 0.333, 0.389$ & 1000 & [9:11]\\
       $20^3 \times 40$ & $0.50$ & $4.0$ & $1.0$ & $0.20, 0.25, 0.30, 0.35$ & 520 & [9:11]\\
       $24^3 \times 48$ & $0.417$ & $4.0$ & $1.0$ & $0.167, 0.208,
       0.25, 0.291$ & 560 & [11:13]\\
    \end{tabular}  
  \end{center}
  \caption{Simulation parameters. Lattice size, bare fermion mass, smearing parameter, gauge fixing parameter, photon mass, the number of configurations, and the 
  points used in the $t \to 0$ extrapolation are indicated.}\label{tab:lattice}
\end{table}

The configurations are generated according to the probability
distribution $\propto e^{-S}$ with the action in Eq.~\eqref{eq:action}. In the
momentum space, the action is diagonalized as 
\begin{align}
  S = {1 \over 2 V} \sum_{k,\mu,\nu} 
  \tilde A_\mu(k) {\cal D}_{\mu\nu}^{-1} (k) \tilde A_\nu (-k),
\end{align}
where ${\cal D}_{\mu \nu} (k)$ is defined in
Eq.~\eqref{eq:photon_propagator} and $\tilde A_\mu(-k) = \tilde
A_\mu^* (k)$. By choosing $\xi = 1$, ${\cal D}^{-1}_{\mu \nu} (k)
\propto \delta_{\mu \nu}$, and thus each component of $\tilde A_\mu
(k)$ has the gaussian distribution with the variance determined by
$k^2$.
 The generation of
the configurations can be done with almost no cost.

By using the gauge configurations, we perform the computations of the
three point function in Eq.~\eqref{SDeq} for the smallest and the next
smallest photon momenta: $k_\mu = (0,0, 2 \pi/L, 0)$ and $(2 \pi /
L,0, 2 \pi/L, 0)$, and extrapolate to $k^2 =0$. 
The data obtained with each configuration is extrapolated to $k^2=0$ with a linear function.
We therefore evaluate the statistical uncertainty for the data point at $k^2=0$ rather than those at finite $k^2$.\fn{
Due to the smallness of the expansion coefficient in $k^2/m_f^2$ in eq.~\ref{finitemassk},
the error of the $k^2=0$ extrapolation is expected to be small.
The careful study of this systematic uncertainty requires more data point
and will be our future work.}
We show in
Fig.~\ref{fig:gtfunc} the perturbative expansion of the function
$g(t)/2$ in Eq.~\eqref{gfacmethod} for the case of $m_\gamma a =
0.167$ on the $24^3 \times 48$ lattice. 
We define $g(t)=g(0)+g(2) (\alpha/\pi)+g(4) (\alpha/\pi)^2+\cdots$ where the perturbative coefficients $g(0), g(2), g(4), ...$ in the right hand side 
are functions of $t$. Extrapolations to $t \to
\infty$ are done with the function $g(t)/2 = a + b/t$ by using two
points around $t \sim L/2$, {\it i.e.}, $t = 11$ and 13.
(The discussion on the behavior of $g(t)$ and the reason of the choice of this fit function is given in ref.~\cite{Kitano:2021ecc}.) 

\begin{figure}[tbp]
\begin{minipage}{0.5\hsize}
    \includegraphics[width=65mm]{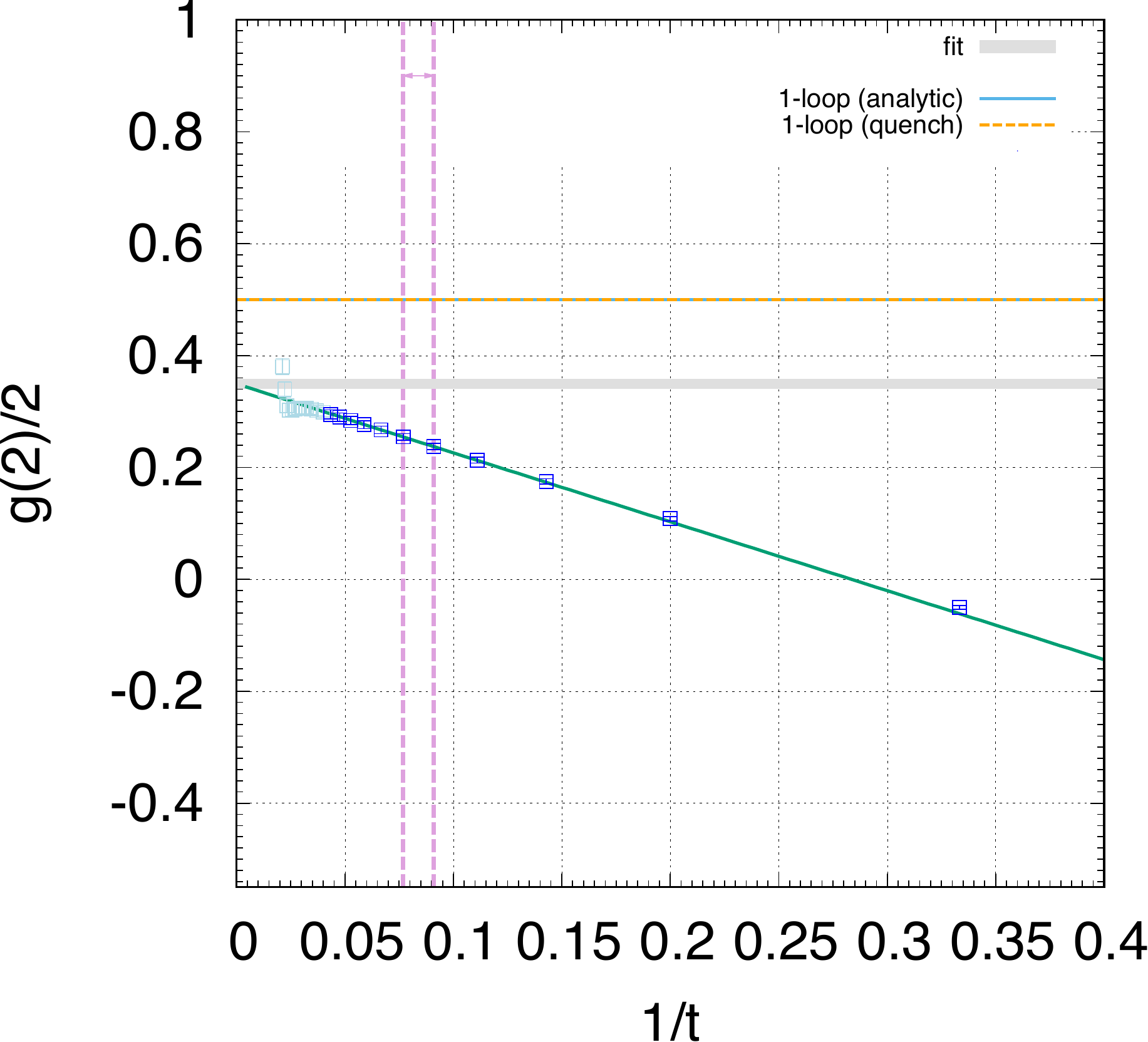}
\end{minipage}
\begin{minipage}{0.5\hsize}
    \includegraphics[width=65mm]{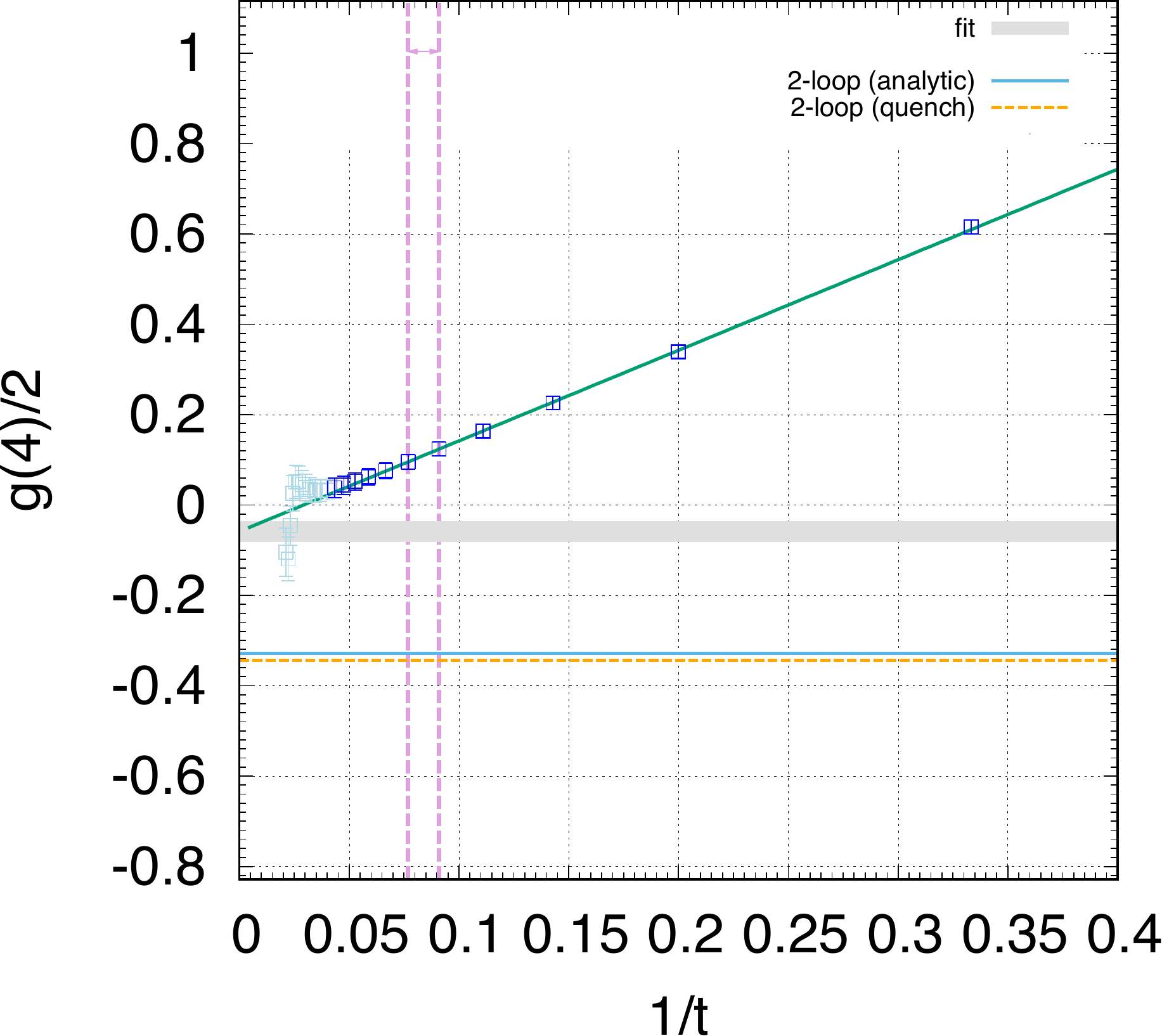}
\end{minipage}

\vspace{1cm}
\begin{minipage}{0.5\hsize}
    \includegraphics[width=65mm]{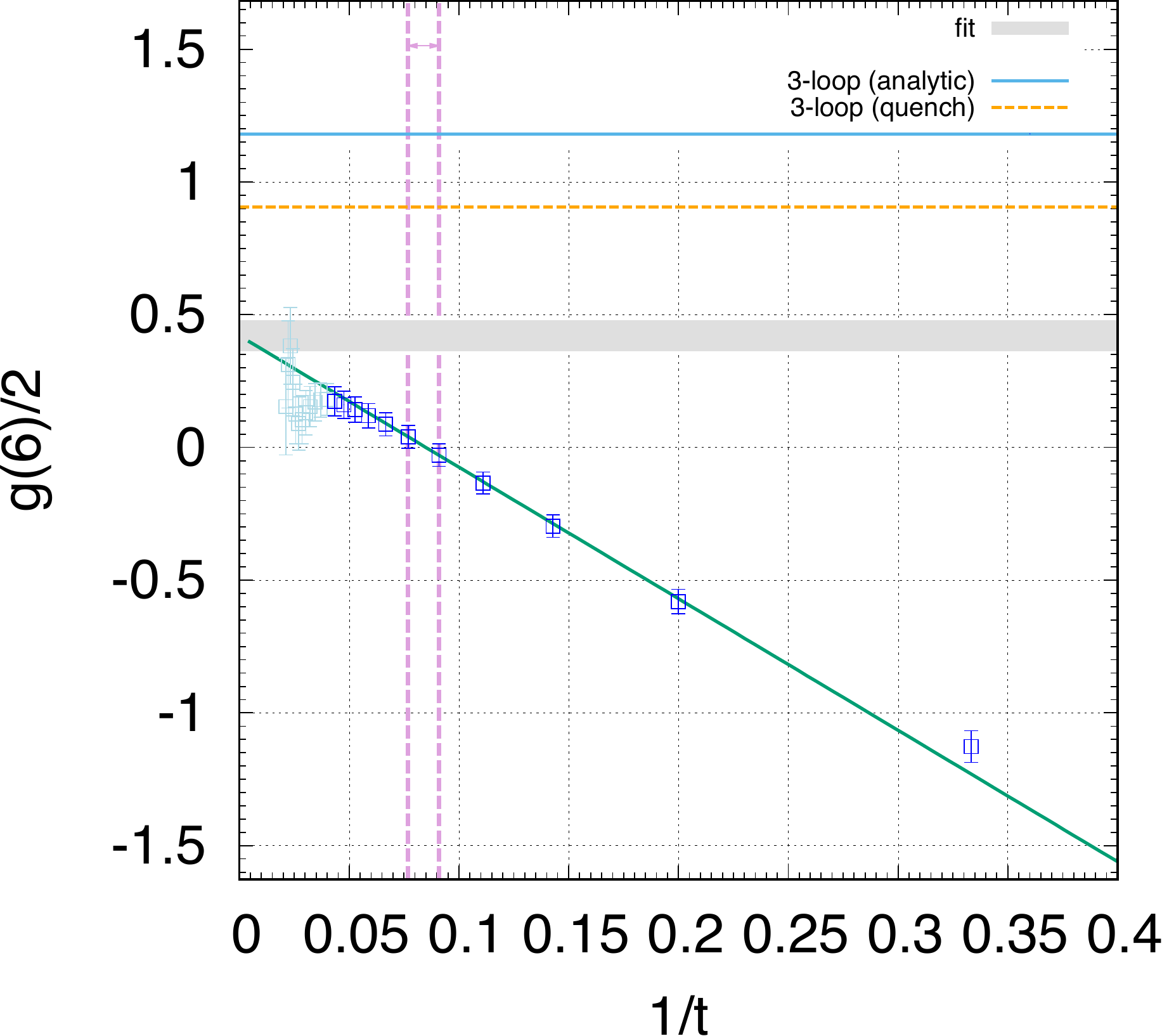}
\end{minipage}
\begin{minipage}{0.5\hsize}
    \includegraphics[width=65mm]{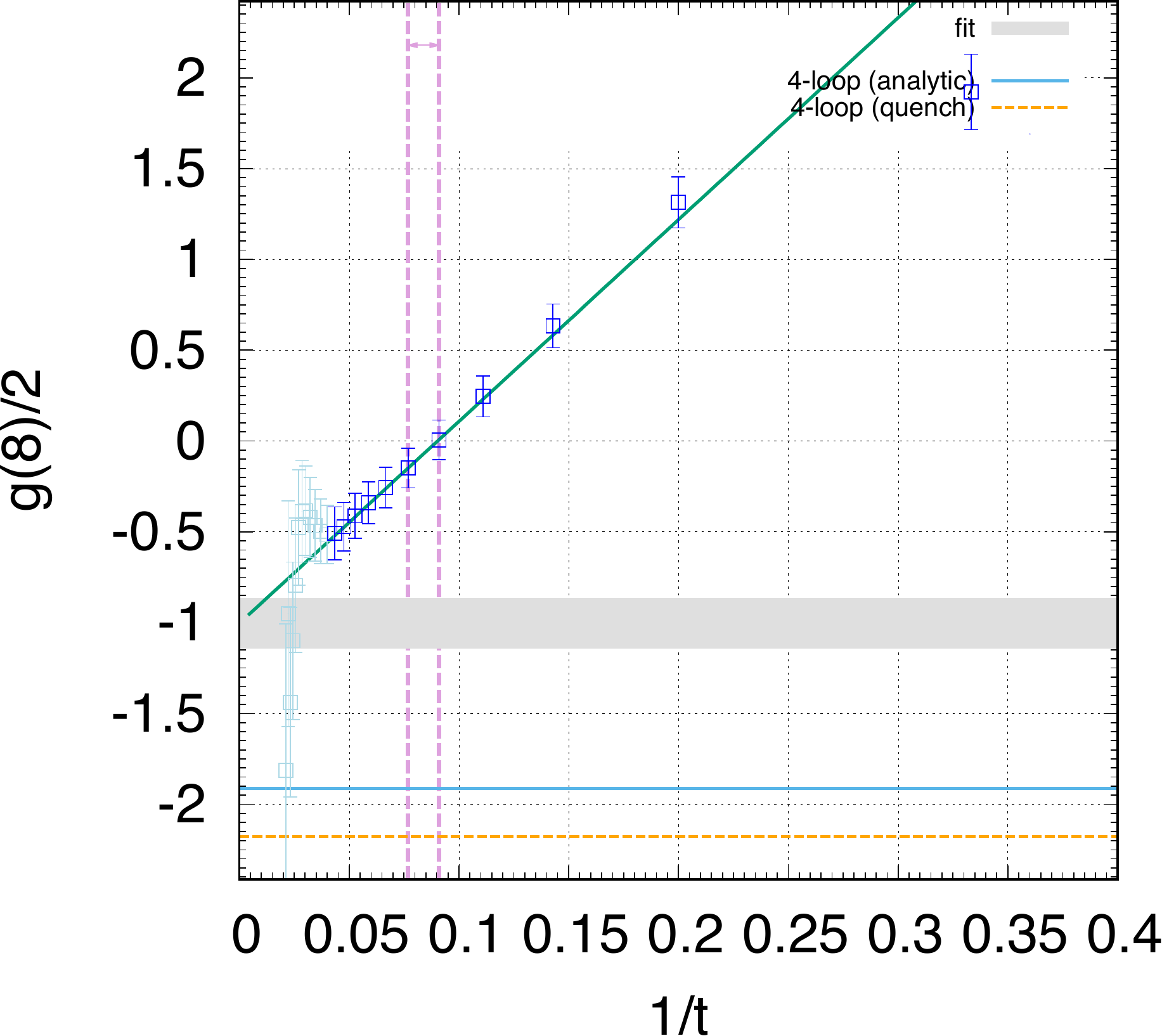}
\end{minipage}

\vspace{1cm}
\begin{minipage}{0.5\hsize}
    \includegraphics[width=62mm]{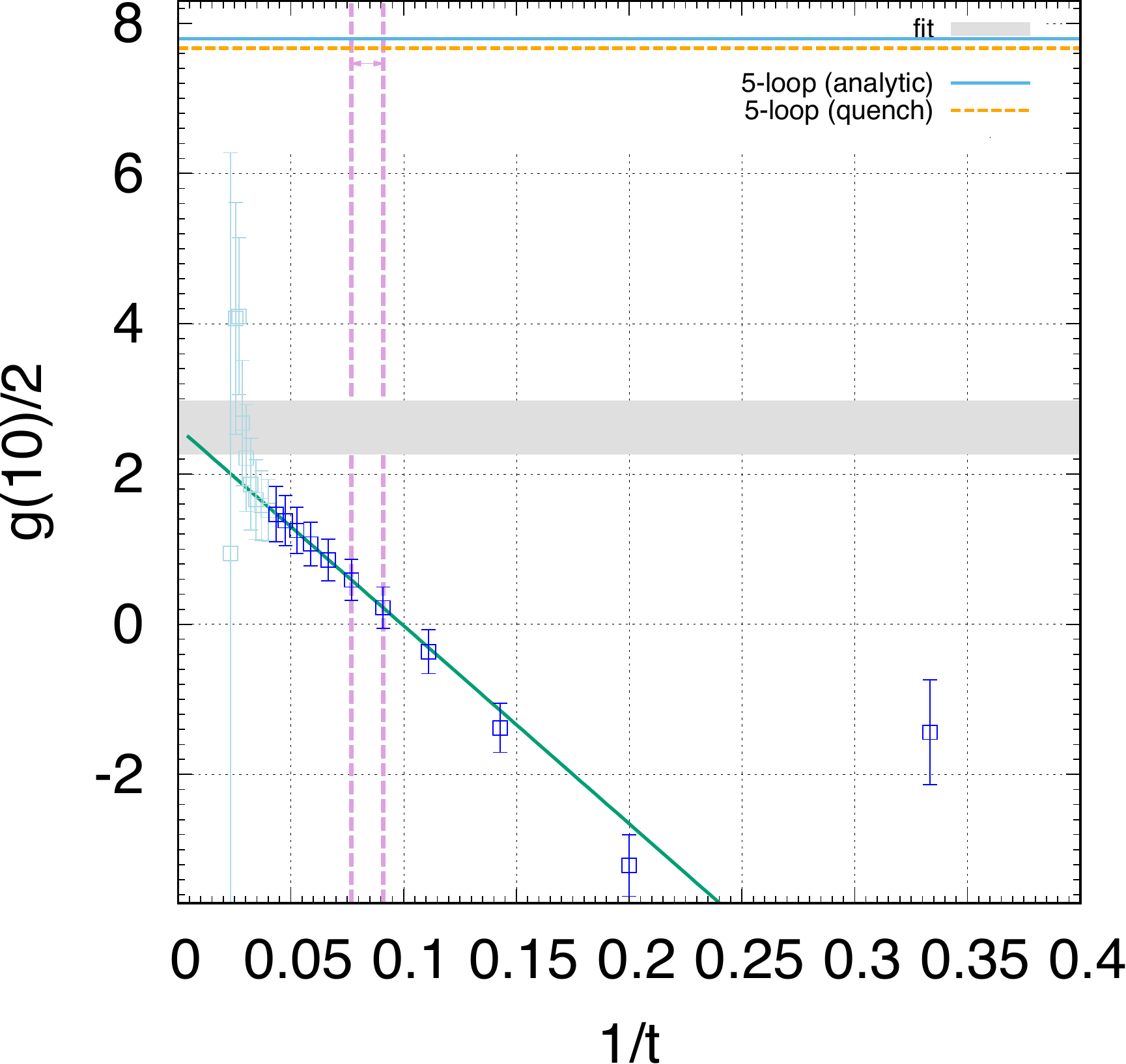}
\end{minipage}
\caption{The perturbative coefficients of the function $g(t)/2$ on
  the $24^3 \times 48$ lattice. The photon mass is $m_\gamma a =
    0.167$. Other parameters are listed in Table~\ref{tab:lattice}. The perturbative coefficients 
    obtained in the standard loop calculations are shown in the case of quenched QED (orange dashed) and QED with the dynamical fermion (blue).
    The two dashed vertical lines show the range of the points used for the fit. The gray band shows the $t=0$ result with its statistical uncertainty.}
\label{fig:gtfunc}
\end{figure}

We examine the stability of the extrapolation to $t \to \infty$ by
comparing with the results with other choices of the two extrapolation
points. In Fig.~\ref{fig:t_ext}, we show the perturbative coefficients
obtained by the extrapolation of the points at $t$ and $t+2$ with the
function $g(t)/2 = a + b/t$. The gray bands represent statistic uncertainties and are obtained with our choice of $t
= 11$. It is expected that for small $t$ the contributions from
excited states, such as a photon-electron two particle state, in the
integrations in Eqs.~\eqref{FE} and \eqref{FM} cannot be ignored.
On the other hand, for large $t$ the finite volume effects
are important. The effects of the backward propagation of the fermion
are enhanced for higher orders in perturbation, as the expansion the
fermion pole mass, $m_f$, in the backward propagation factor, $e^{-
m_f (T-t)}$, gives large coefficients for higher orders. As discussed
in Sec~\ref{sec:3.2}, we used both the periodic and
anti-periodic boundary conditions of fermions to evaluate three-point
functions at the fermion Euclidean energies $p_4 = 0, \pi/T, 2\pi/T,
\cdots$. This treatment effectively enlarges the extent in the time
direction $T$ to $2T$. The backward propagation is significantly
suppressed, but still it is better to be away from a large $t$ region.
The choice of $t \sim L/2$ seems to give reasonably stable values
although it is desired to have larger statistics for the confirmation
of the stability for three loops and higher.

\begin{figure}[tbp]

\begin{minipage}{0.5\hsize}
    \includegraphics[width=65mm]{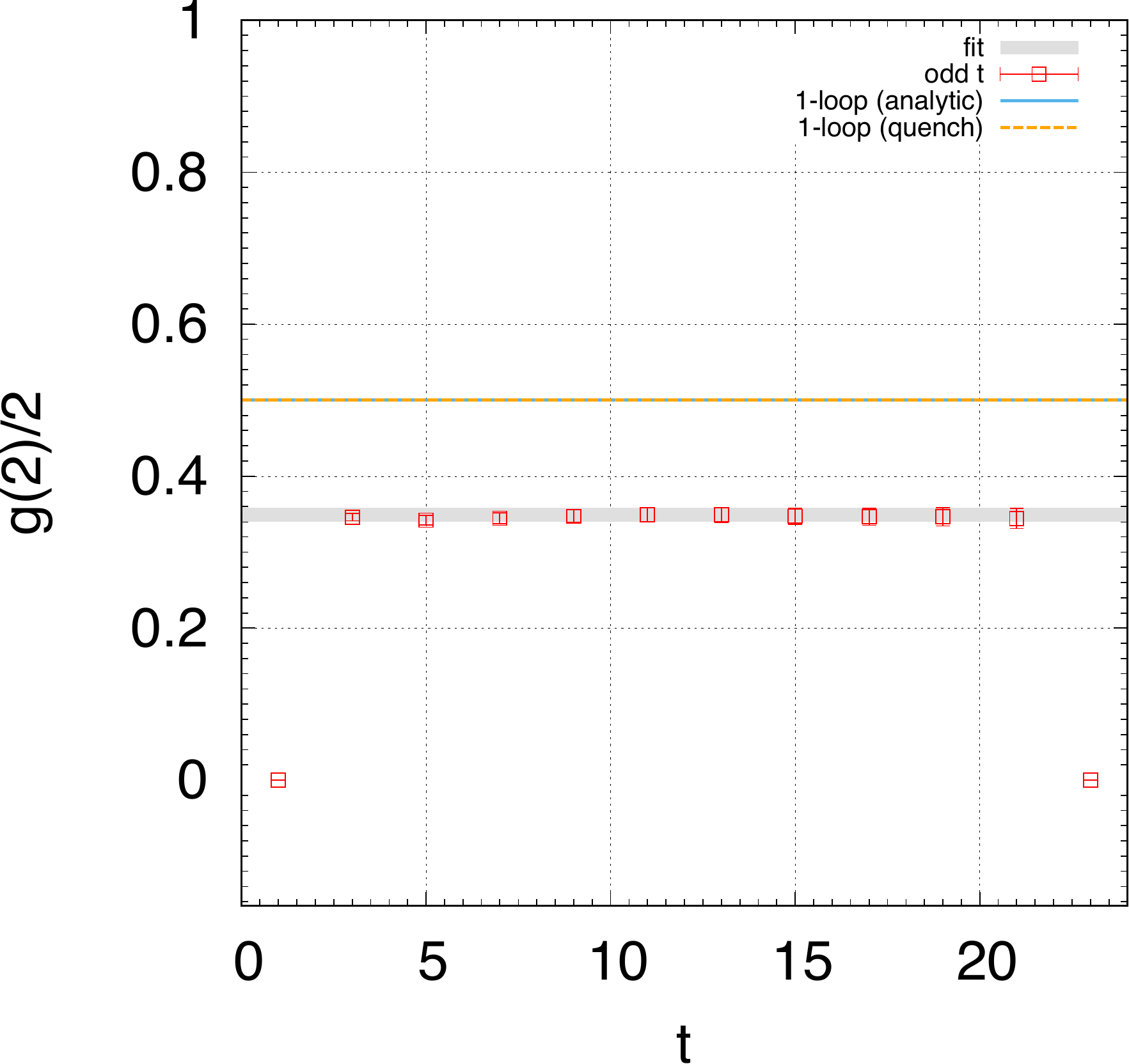}
\end{minipage}
\begin{minipage}{0.5\hsize}
    \includegraphics[width=65mm]{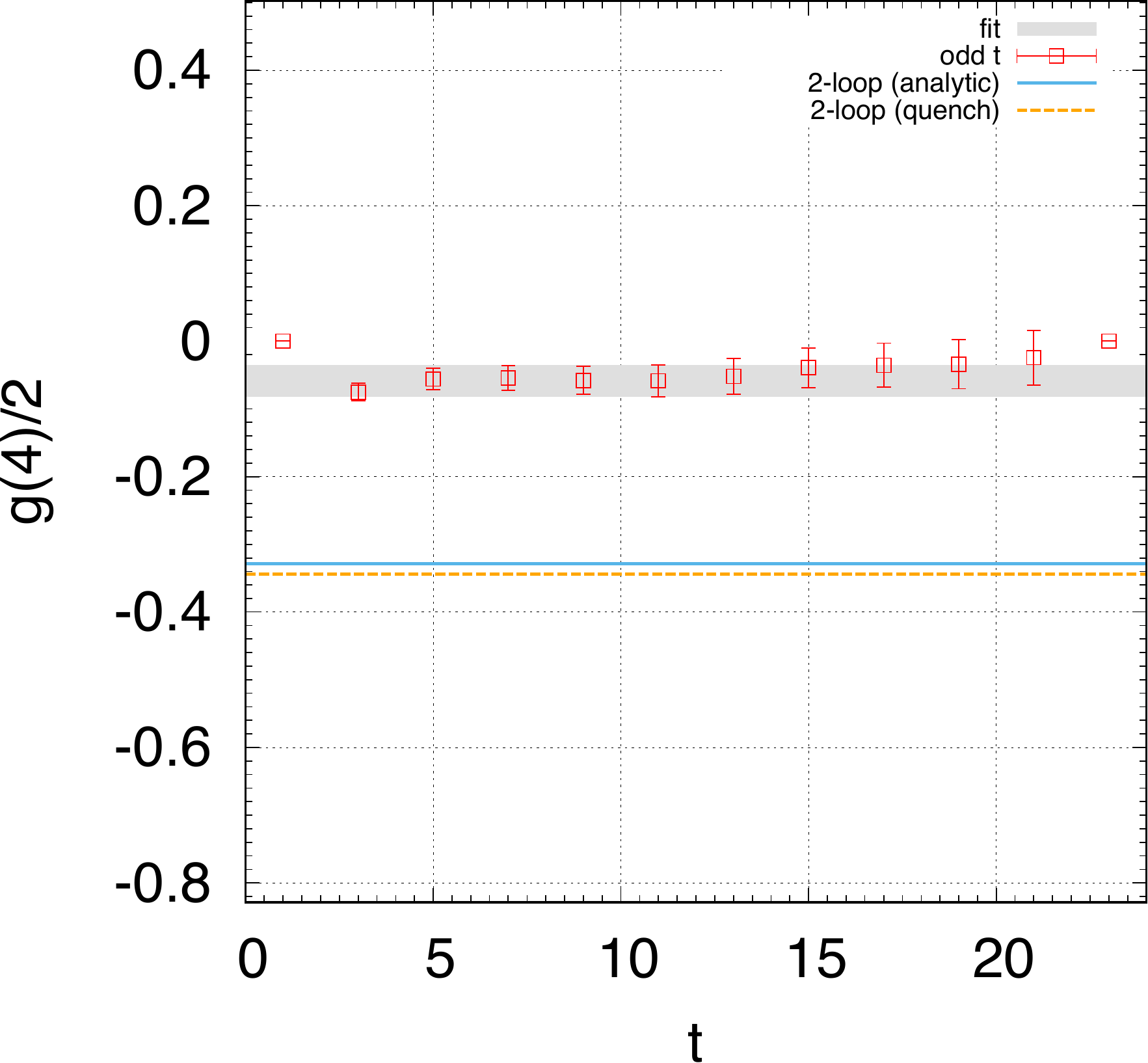}
\end{minipage}

\vspace{1cm}
\begin{minipage}{0.5\hsize}
    \includegraphics[width=65mm]{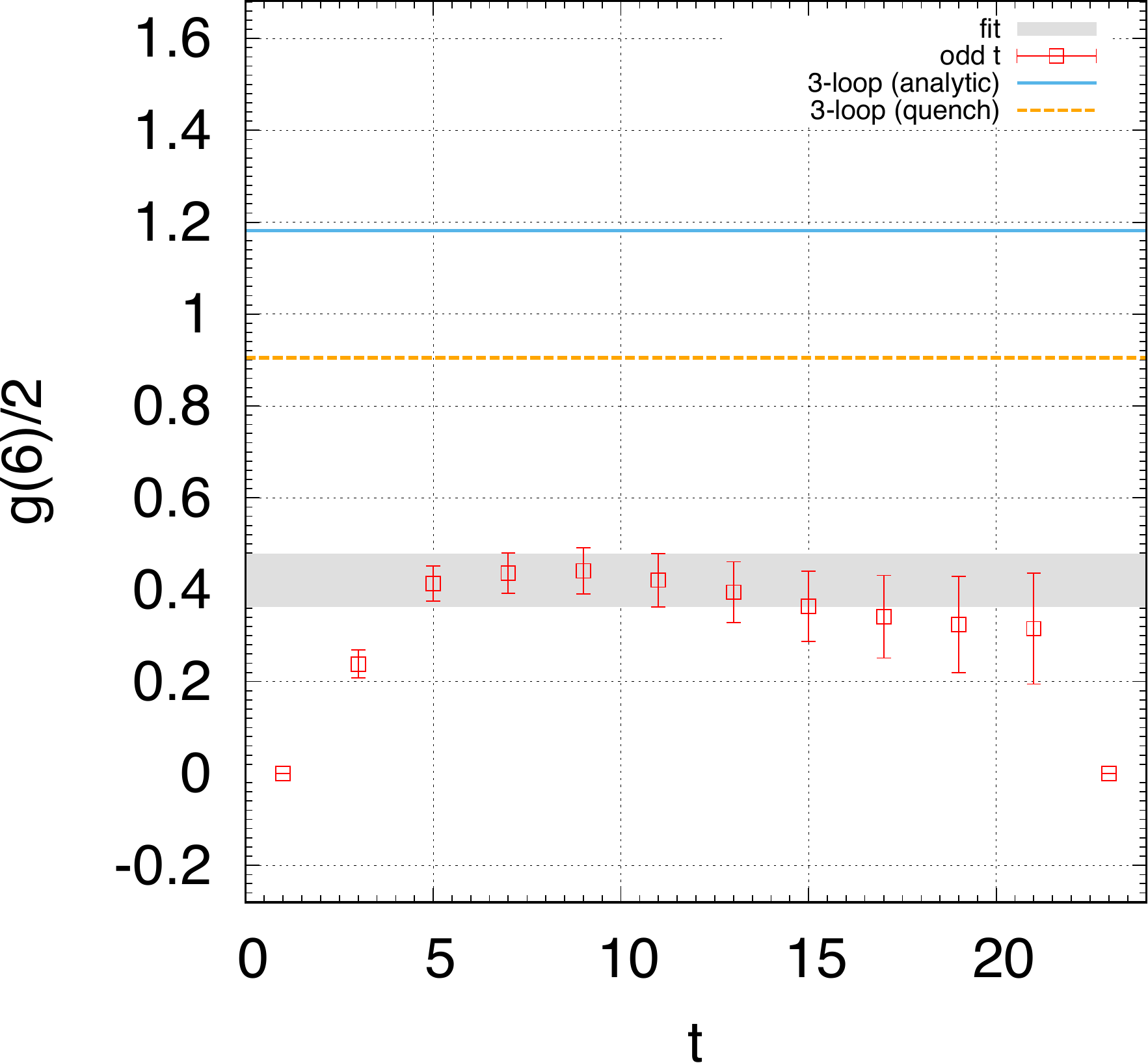}
\end{minipage}
\begin{minipage}{0.5\hsize}
    \includegraphics[width=65mm]{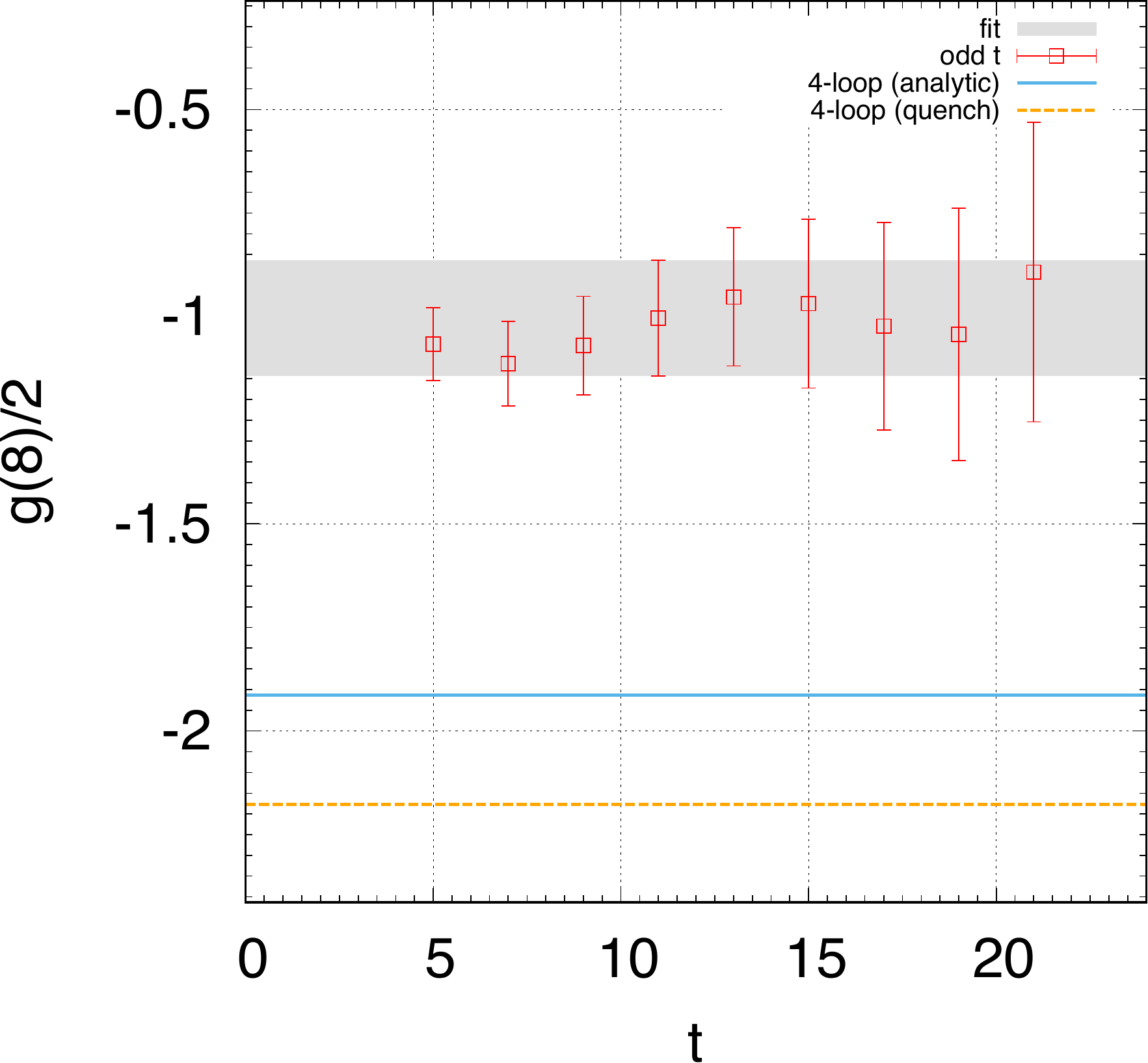}
\end{minipage}

\vspace{1cm}
\begin{minipage}{0.5\hsize}
  \includegraphics[width=62mm]{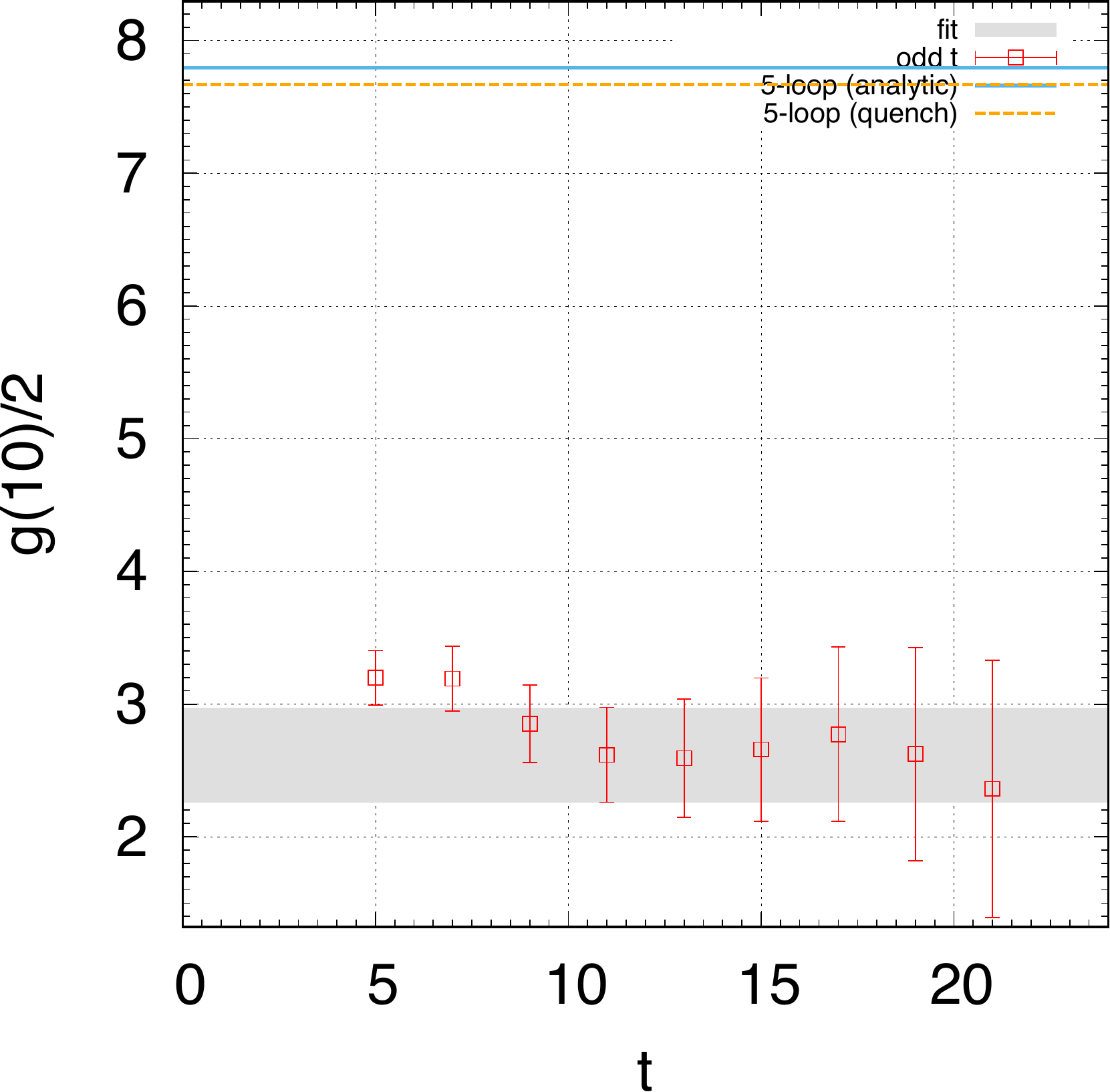}
\end{minipage}

  \caption{The $t=0$ results obtained by extrapolating the points at $t$ (horizontal axis) and $t+2$ on
  the $24^3 \times 48$ lattice.
  The gray bands, which are the results obtained by extrapolating the points at $t=11$ and $13$, 
  are also shown to examine the validity of them.}
  \label{fig:t_ext}
\end{figure}

\begin{figure}[tbp]
  \begin{minipage}{0.5\hsize}
    \includegraphics[width=65mm]{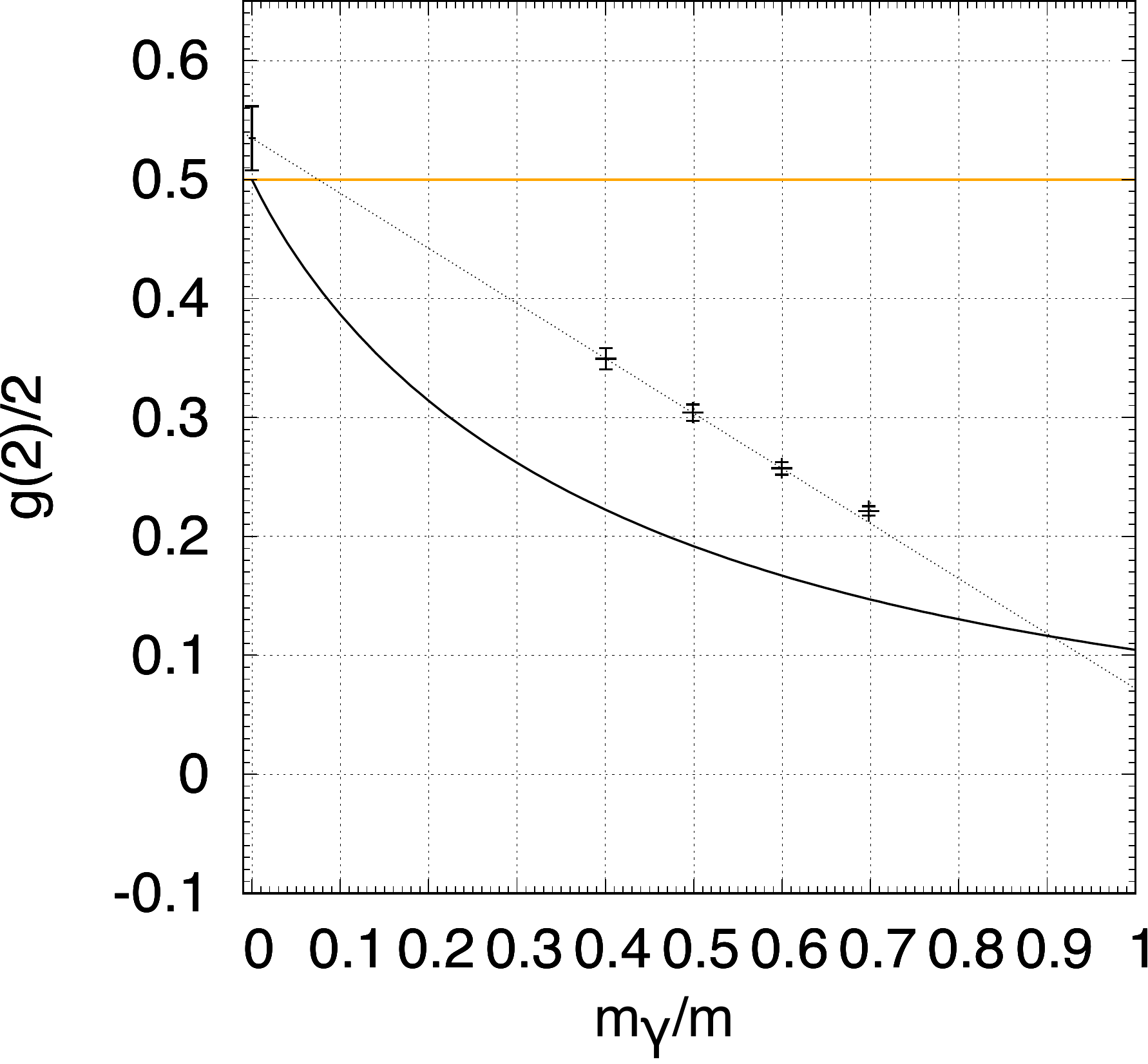}
  \end{minipage}
  \begin{minipage}{0.5\hsize}
    \includegraphics[width=65mm]{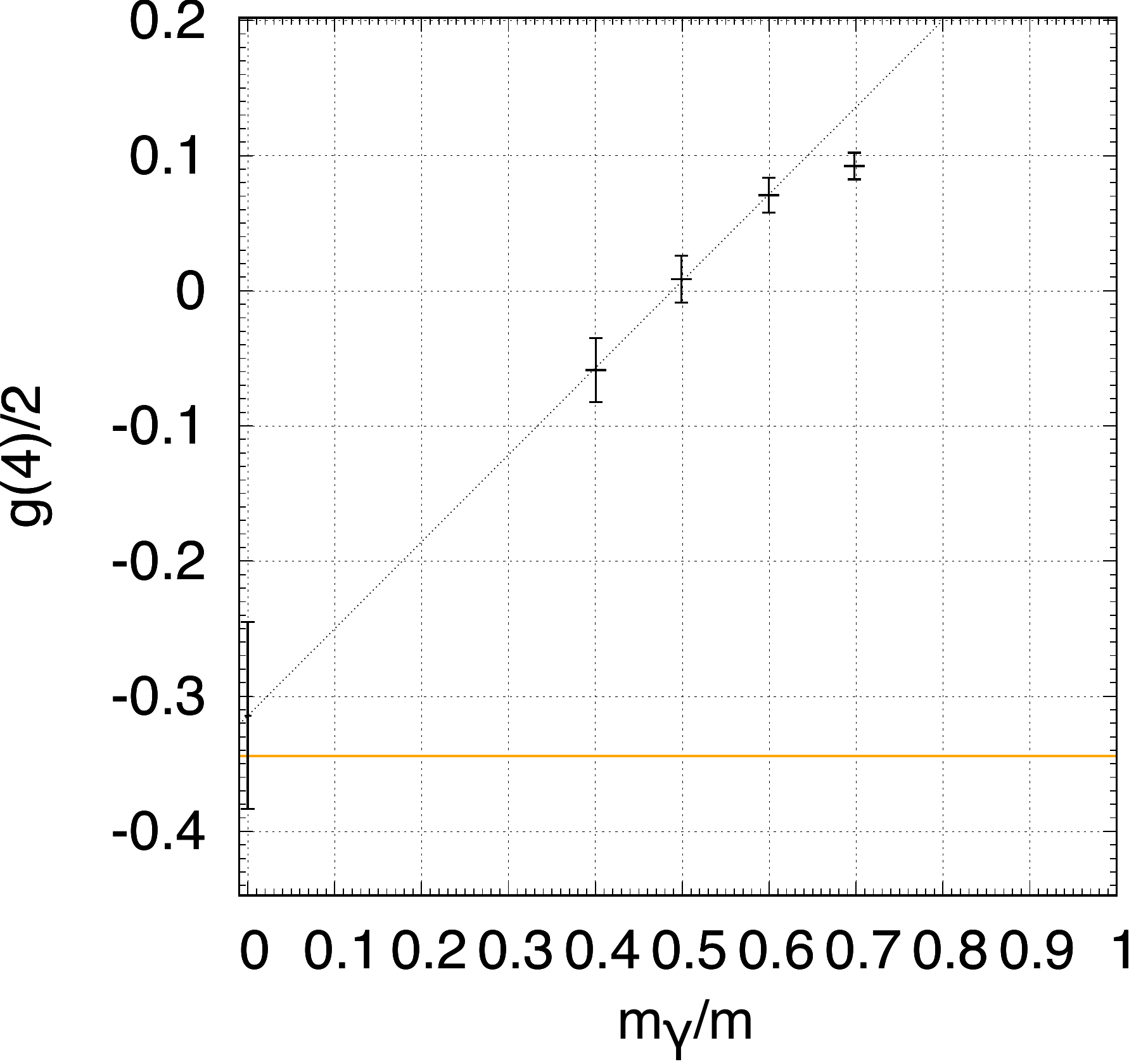}
  \end{minipage}

  \vspace{1cm}
  \begin{minipage}{0.5\hsize}
    \includegraphics[width=65mm]{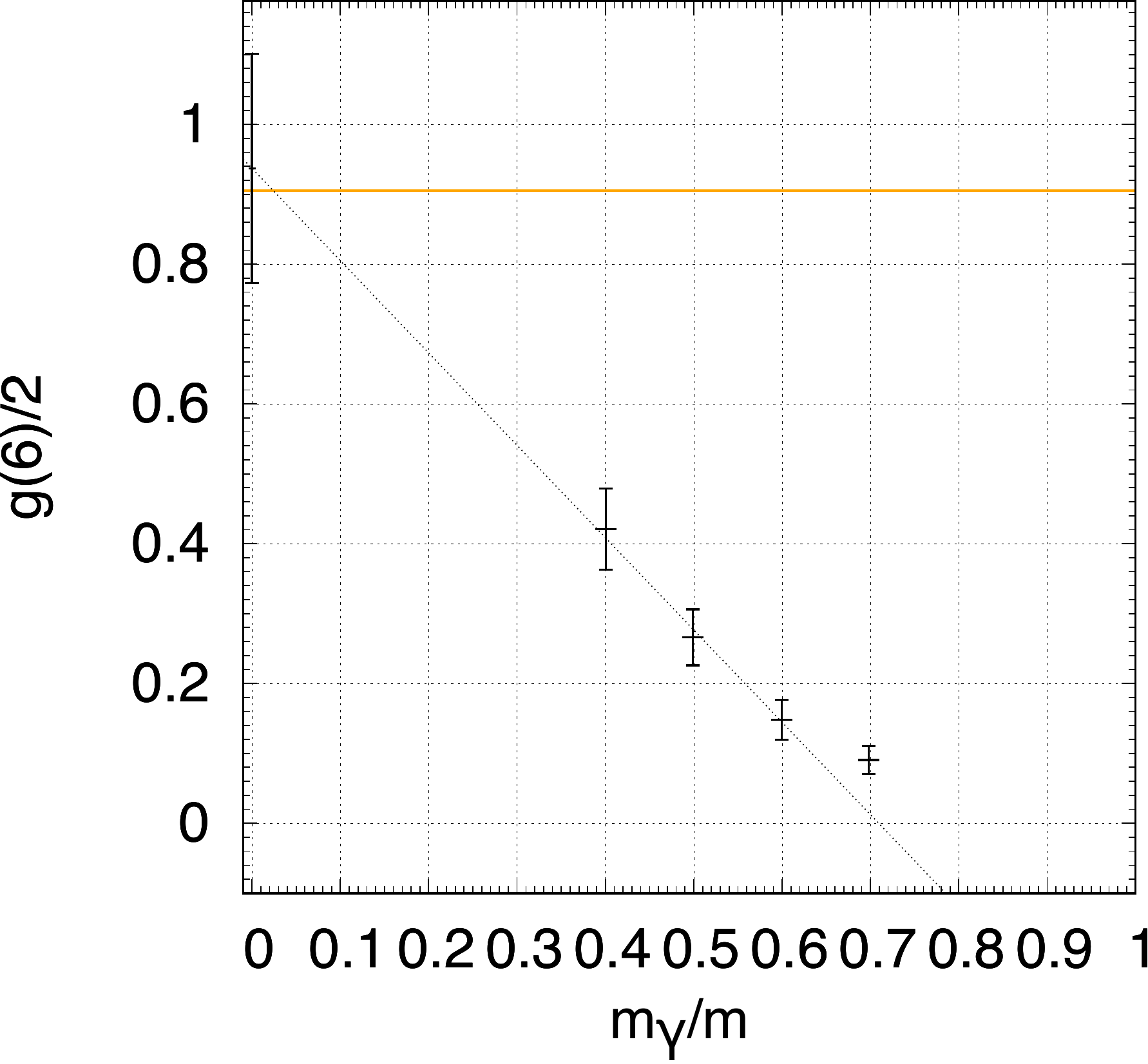}
  \end{minipage}
  \begin{minipage}{0.5\hsize}
    \includegraphics[width=65mm]{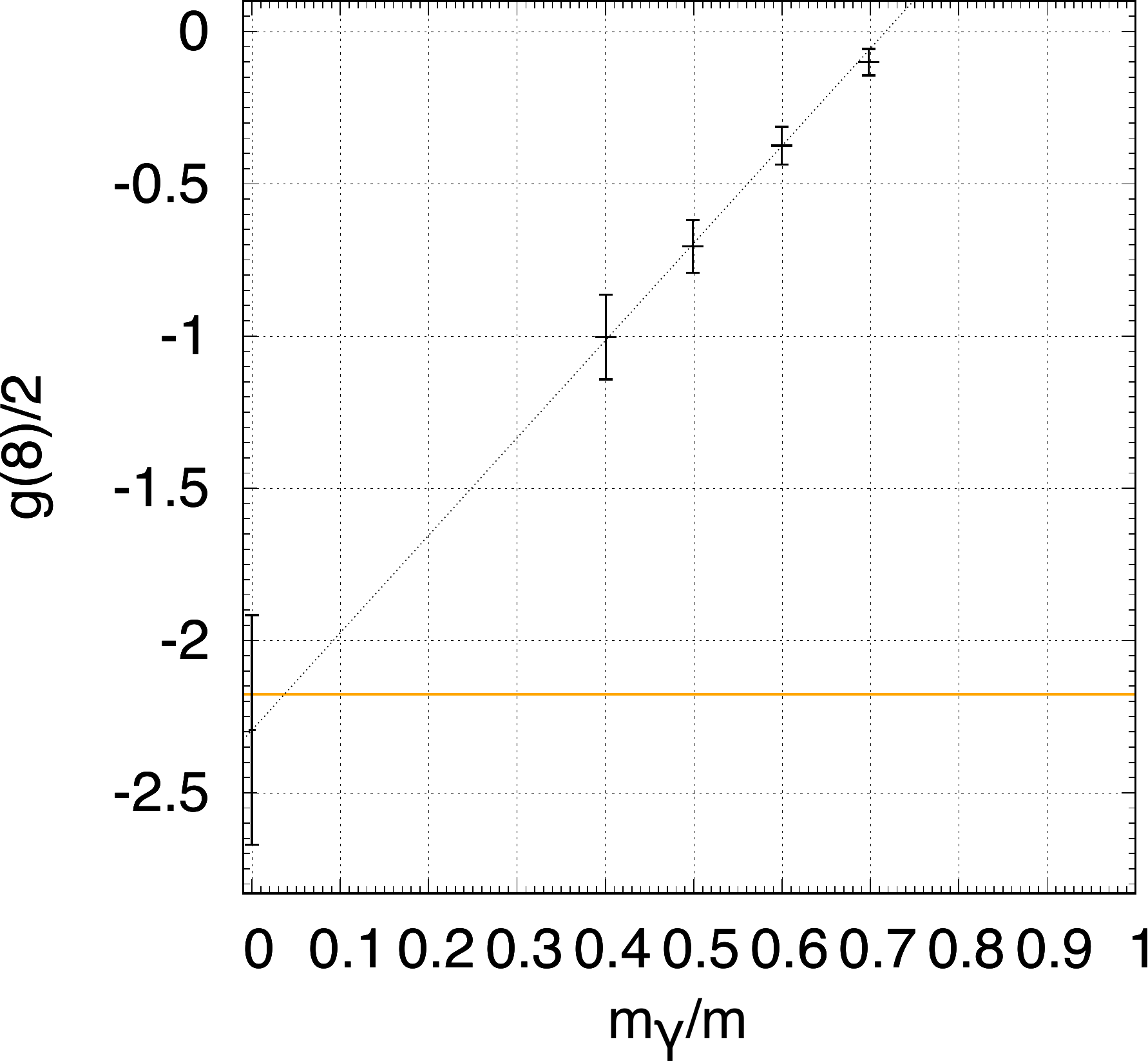}
  \end{minipage}

  \vspace{1cm}
  \begin{minipage}{0.5\hsize}
    \includegraphics[width=62mm]{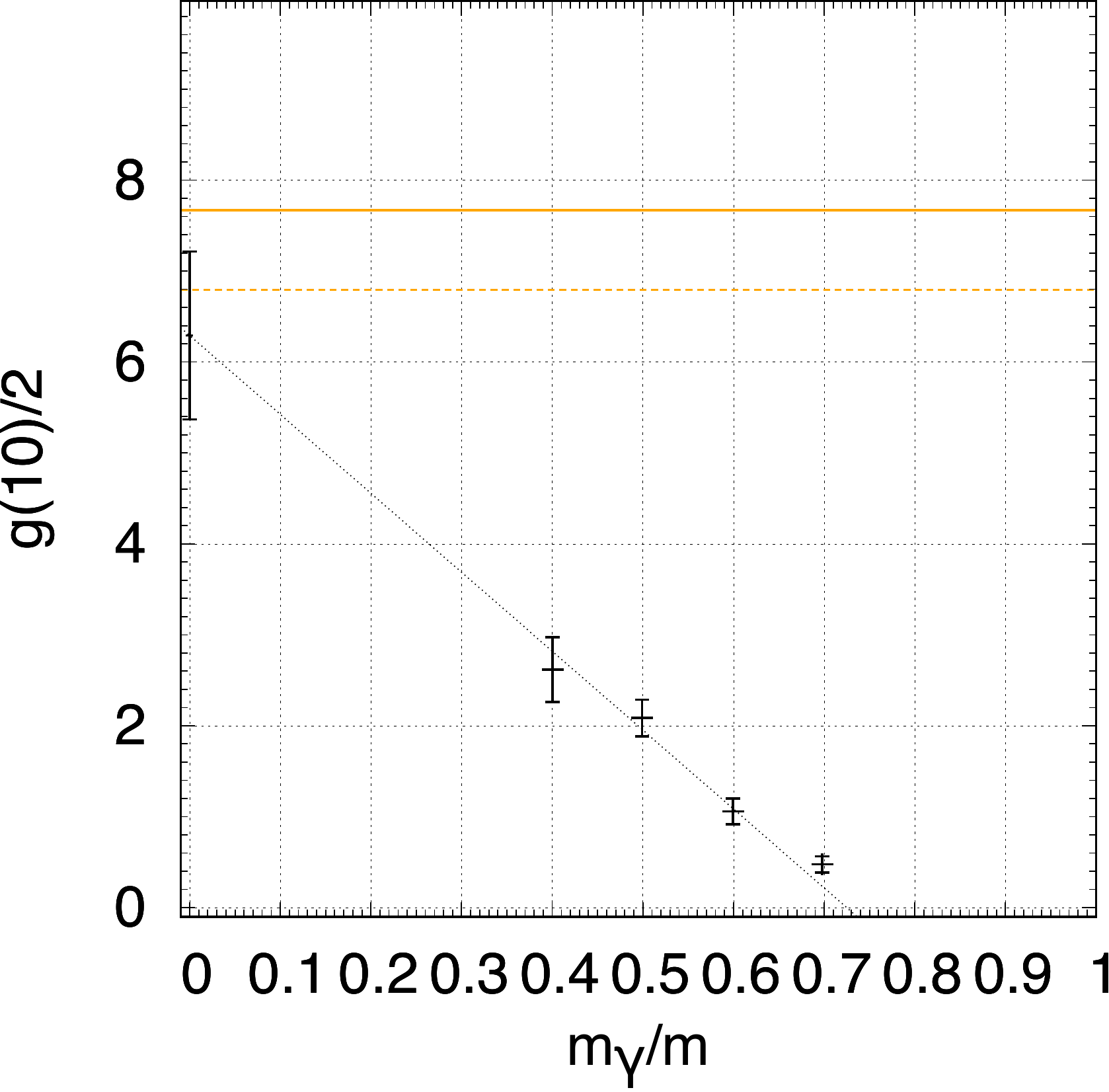}
  \end{minipage}

  \caption{Extrapolation to $m_\gamma = 0$ on the $24^3 \times 48$ lattice. 
  Horizontal lines show the results from Feynman diagram computations (at $k^2=0$ and $m_{\gamma}=0$). For the five-loop coefficient,
  the solid line shows the result of refs.~\cite{Aoyama:2017uqe, Aoyama:2019ryr} while
  the dashed one the result of ref.~\cite{Volkov:2019phy}.
  In the first figure, the solid curve represents the finite photon mass correction theoretically calculated in eq.~\eqref{finitemassk}
  (while setting $k^2=0$) in continuum spacetime.}
  \label{fig:mg_ext}
\end{figure}

Obtained values of $g(t \to \infty)/2$ are plotted for four choices of $m_\gamma a$ in
Fig.~\ref{fig:mg_ext}. Three points $m_\gamma/m = 0.4, 0.5, 0.6$ are
used for the linear extrapolation to $m_\gamma = 0$ while fixing $m$.
Expected systematic uncertainties
by this extrapolation, 
are of order 20--30\% 
due to higher order terms in the 
$m_\gamma$ expansion, i.e., 
$\mathcal{O}(m_{\gamma}^2/m_f^2)$. 
The effects of the 
discretization are expected to be larger for higher
orders in perturbations. In addition to the discretization effects of
$O(m^2 a^2)$, the UV cut-off in the photon kinetic term gives
corrections typically,
\begin{align}
  e^{- 2 m^2/\Lambda^2},
\end{align}
as a multiplicative factor for each photon propagator. For a choice of
too small $\Lambda / m$, we obtain exponentially suppressed values.
The choice of $\Lambda^2 = 4.0$ seems to be not too small. Further
large values, however, make the statistical error larger and also the
logarithmic correction larger. One should look for optimal values of
$\Lambda$ for different choices of simulation parameters.

We show the analytic results of $g/2$ with finite photon masses at the
one-loop level (using the first 
equality of eq.~\eqref{finitemassk} with $k^2=0$)
in Fig.~\ref{fig:mg_ext} 
as the solid curve. The
lattice results are significantly off the line, which represents the
discretization effects. With the finite photon mass, there can be
discretization errors of order $m_\gamma a$ in addition to $m^2 a^2$.
Therefore, the limit of $m_\gamma \to 0$ would give the closest value
to the continuum theory. Such a tendency can be seen in the figure.

We repeat the same analyses for $L=20$, 18, 16 and 14, while keeping
$m_\gamma/m = 0.4-0.7$ and $m_\gamma L = 4$ for the
smallest $m_\gamma$ as listed in Table~\ref{tab:lattice}.
For smaller lattices, the fermion mass $ma$ is larger, and thus results are
further from the continuum limit. We show in Fig.~\ref{fig:cont_limit}
the $m^2 a^2$ dependence of the $g$ factor on each lattice. We see the
tendency of approaching towards the correct values.\footnote{
For $g(8)$ and $g(10)$, the data points are not fitted well in fig.~\ref{fig:cont_limit}.
We consider that this is due to the large systematic error in the $m_{\gamma} \to 0$ extrapolation,
as $m_{\gamma}$ cannot be taken small enough in this study.
Note that the data points only have statistical uncertainties.
}

We refrain from presenting the values of our result
because they are not conclusive ones; 
as we noted, the photon mass cannot be taken small enough in the present study,
and the extrapolation to $m_{\gamma} \to 0$ can have large errors.
For instance, the data point at $m_{\gamma}/m=0.4$ for $g(2)$ 
has an error of $\sim 40\,\%$, as seen from the first figure in fig.~\ref{fig:mg_ext}.
The detailed study of systematic uncertainties is skipped in this work accordingly.

Although the statistical uncertainties are large, the calculation up
to the five-loop level seems to be doable in a larger scale
simulation.
For $L=128$, for example, one can take $m_\gamma / m = 0.1$ and $m^2
a^2 = 0.1$ while satisfying $m_\gamma L = 4$. This would significantly
reduce the uncertainties of the $m_\gamma \to 0$ extrapolation.


\begin{figure}[tbp]

\begin{minipage}{0.5\hsize}
  \includegraphics[width=65mm]{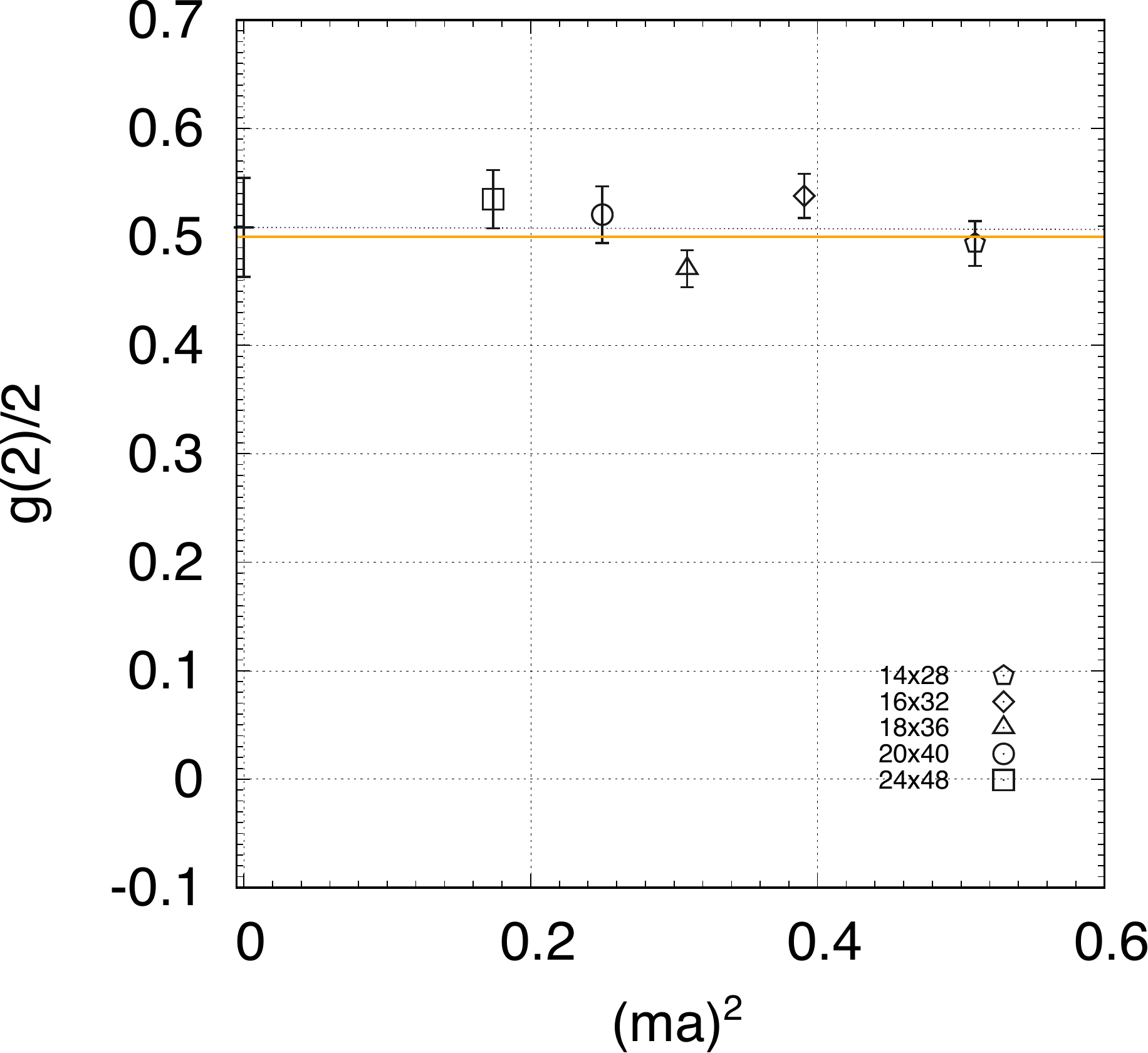}
\end{minipage}
\begin{minipage}{0.5\hsize}
  \includegraphics[width=65mm]{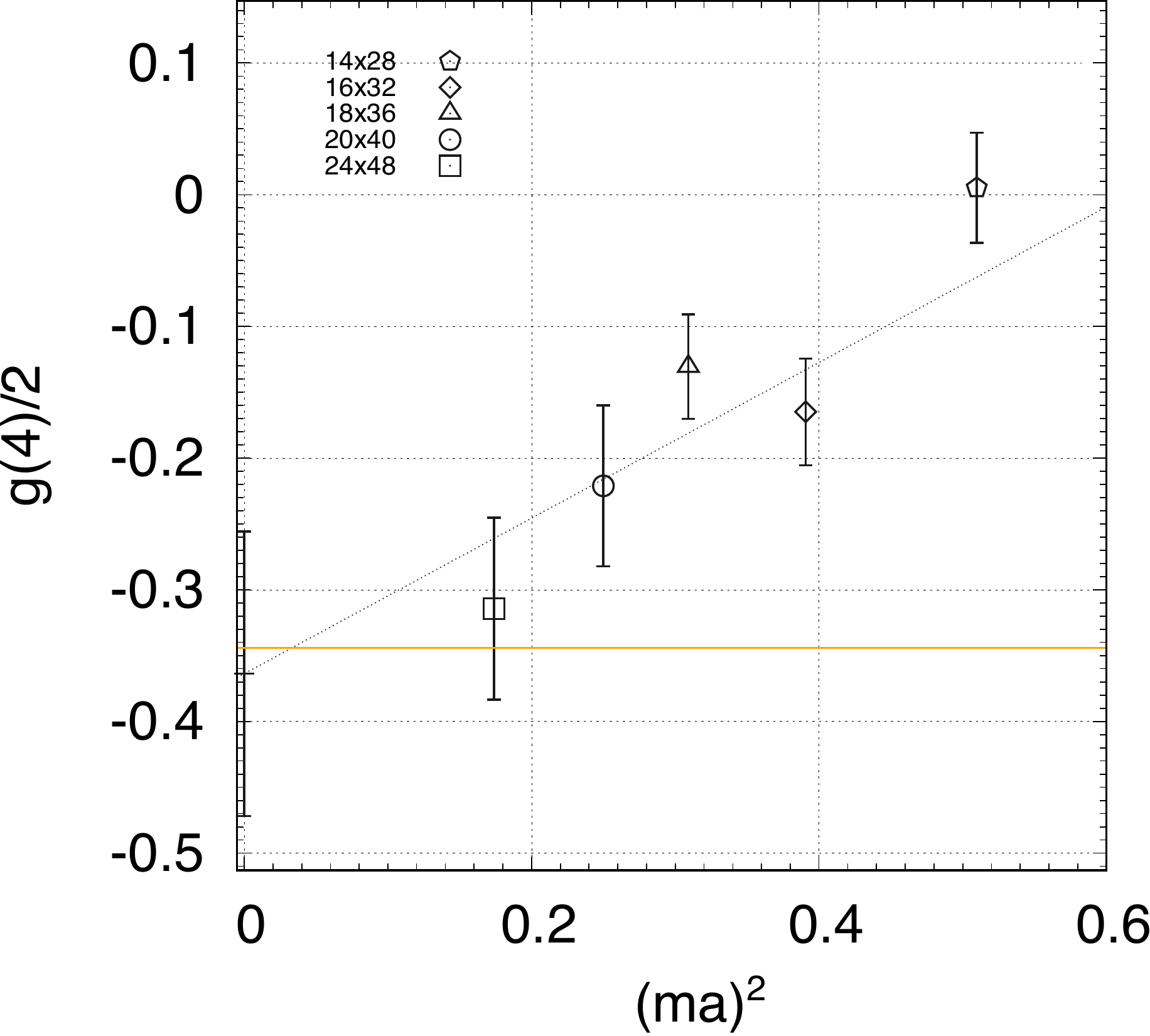}
\end{minipage}

\vspace{1cm}
\begin{minipage}{0.5\hsize}
  \includegraphics[width=65mm]{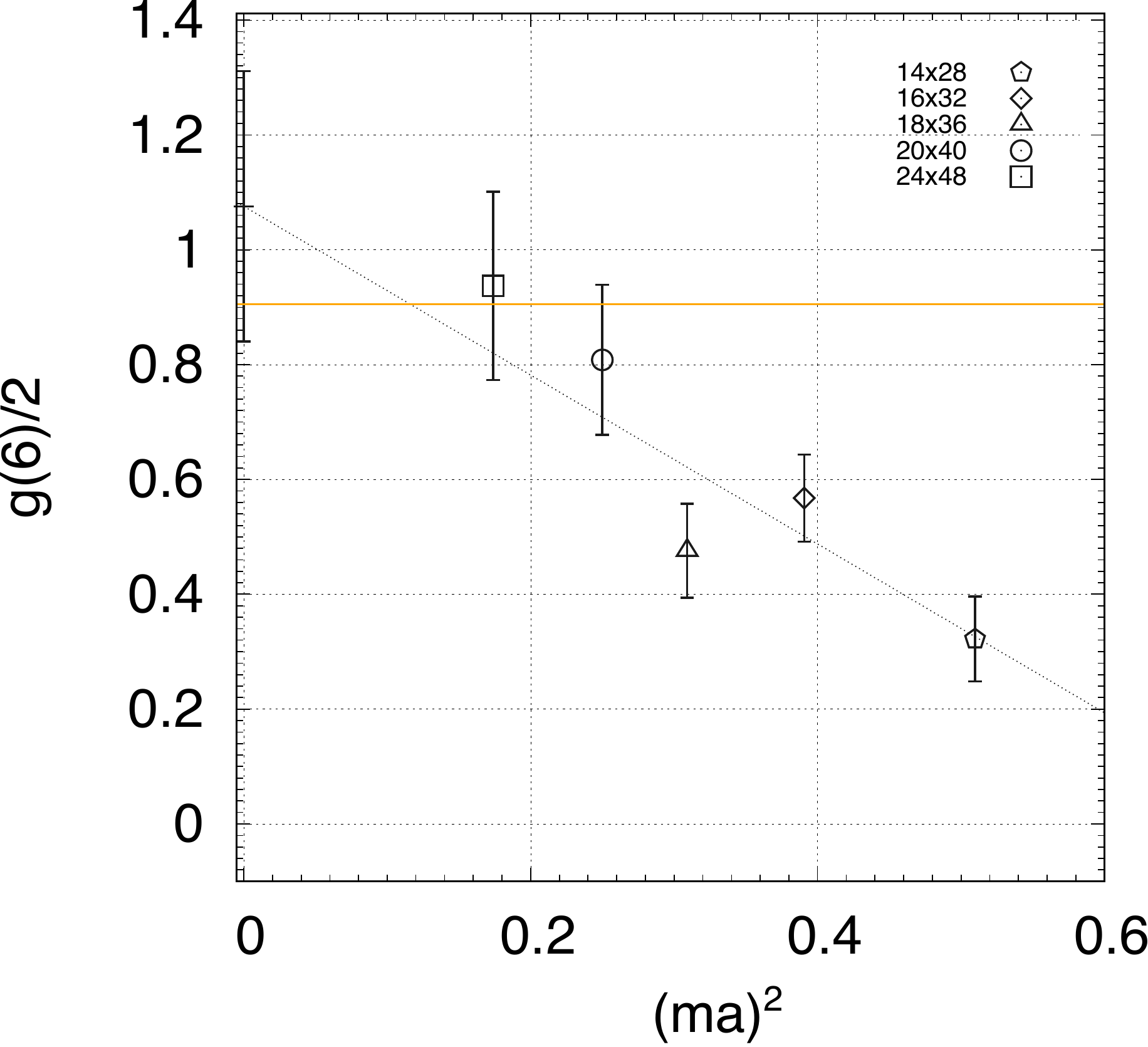}
\end{minipage}
\begin{minipage}{0.5\hsize}
  \includegraphics[width=65mm]{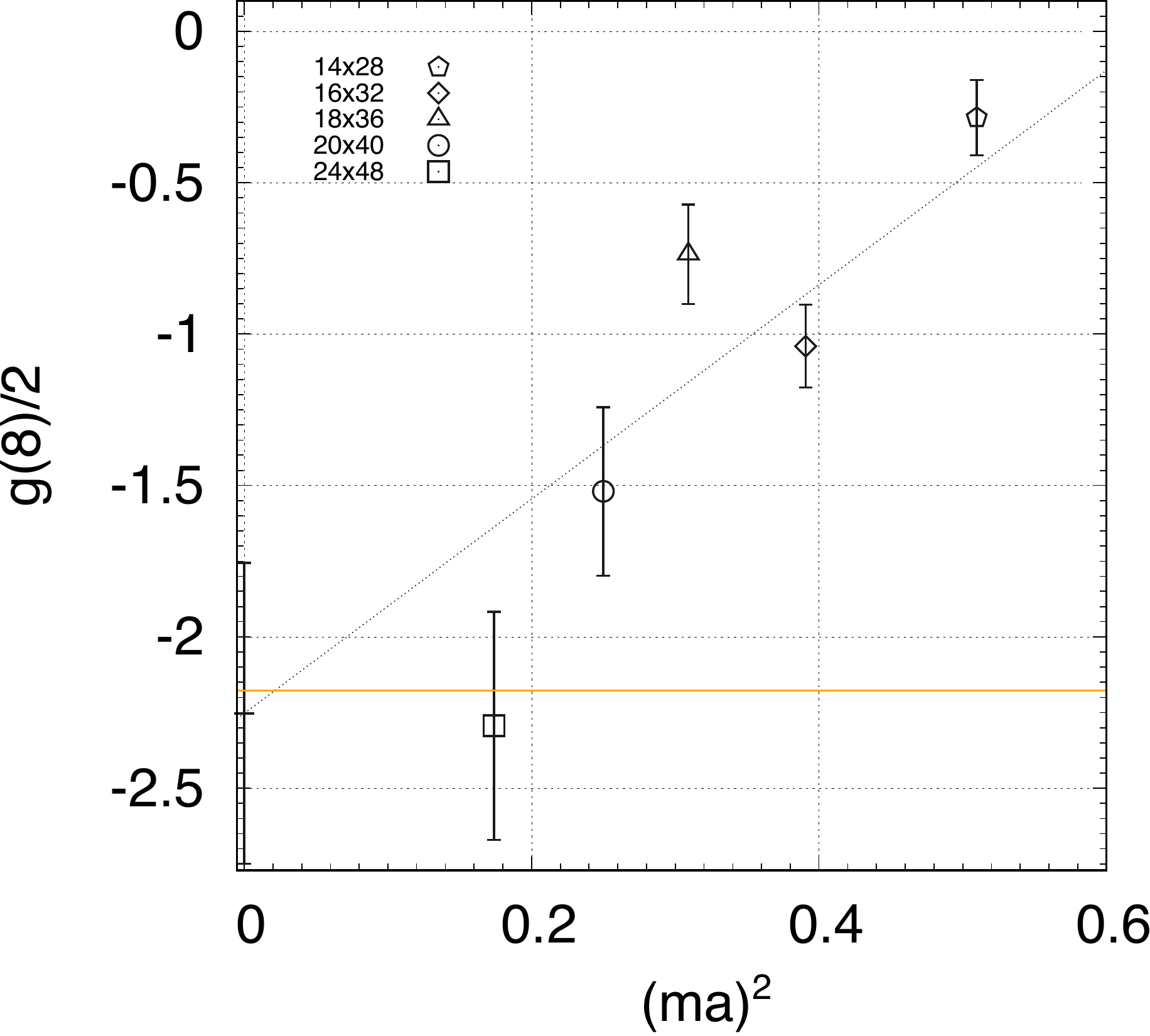}
\end{minipage}

\vspace{1cm}
\begin{minipage}{0.5\hsize}
  \includegraphics[width=62mm]{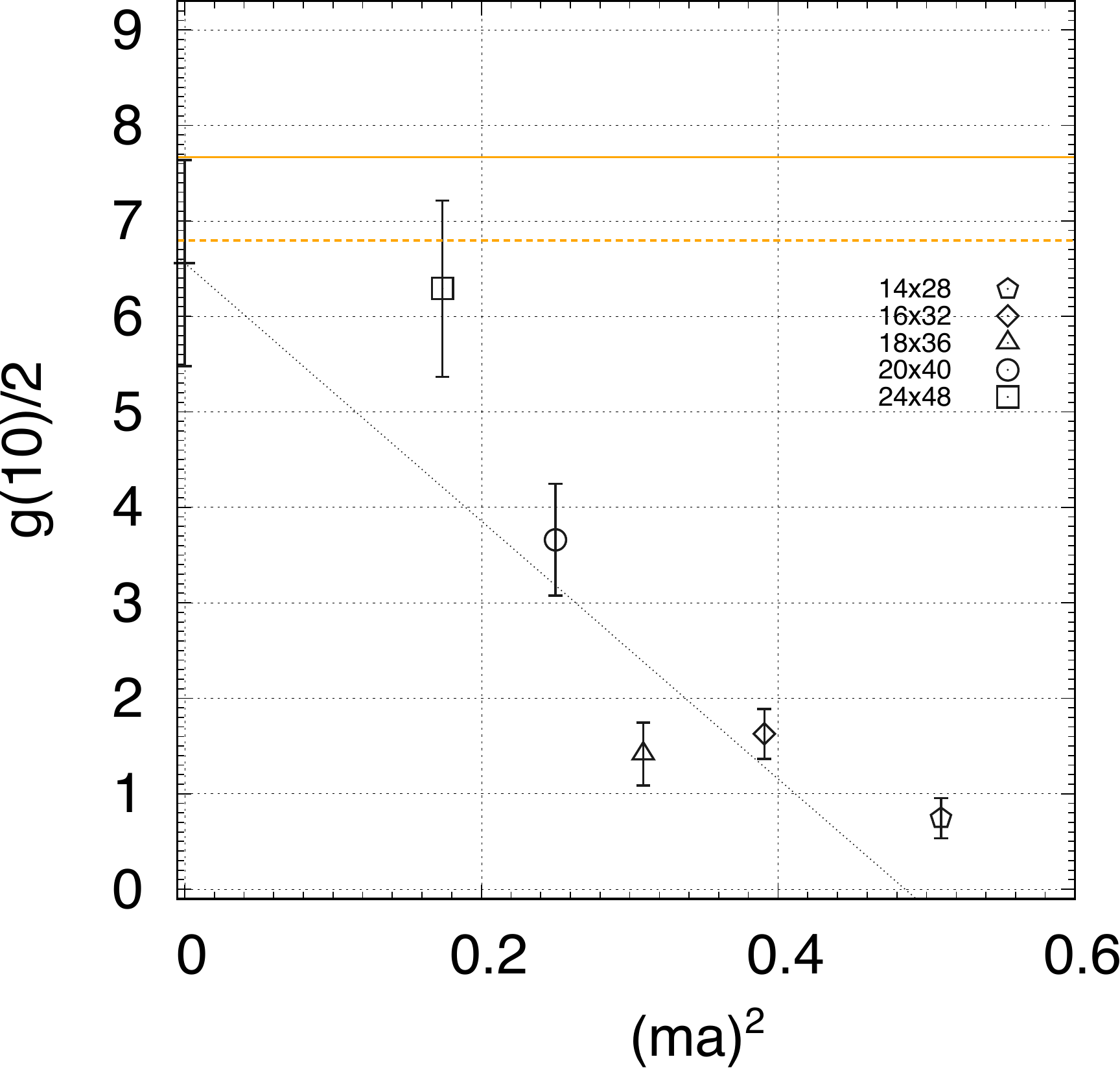}
\end{minipage}

\caption{The continuum limit. For the five-loop coefficient,
  the solid line shows the result of refs.~\cite{Aoyama:2017uqe, Aoyama:2019ryr} while
  the dashed one the result of ref.~\cite{Volkov:2019phy}.}
  \label{fig:cont_limit}
\end{figure}

\section{Conclusions and discussion}
\label{sec:4}

In this paper, aiming at giving the 
five-loop coefficient of the
electron $g$ factor using the lattice,
we developed a theoretical study of
finite volume corrections and
also performed a numerical
simulation using the small lattices.
We first studied 
finite volume corrections in various
IR regularization methods of lattice
QED to discuss optimal regularization for our purpose.
We found that in QED$_L$ finite volume corrections 
to the effective mass can have different
parametric dependences depending on the size
of Euclidean time $t$.
The `naive on-shell result' turns out to be valid only
for $t \gg L$ . 
We also discussed a possible method to
make the FV correction smaller than $\sim 1/(mL)$.
In contrast to such complexity in QED$_L$,
finite photon mass regularization
seems to always suppress finite
volume corrections exponentially
$\sim e^{-m_{\gamma} L}$.
We therefore adopt finite photon
mass regularization.

In addition to finite volume
effects, we studied corrections to
the $g$ factor due to finite
photon mass and finite photon
momentum $k^2$. We gave the parametric
dependences on these perturbations;
this understanding is used to
fix extrapolation functions. 
Based on these studies on systematic
errors, we presented an optimal
strategy for selecting the simulation
parameters $m_{\gamma}, m,
\Lambda_{\rm UV}^2, L$ and the 
order of various extrapolations.

We presented a numerical lattice simulation following our strategy. 
Due to the limited lattice volume in our study, the evaluation of the systematic uncertainties, such as those in the $m_\gamma \to 0$ and $a \to 0$ extrapolations, need more study.
With our choice of parameters, we observed large discretization
effects. This can be understood from discretization effects of
$\mathcal{O}(m_{\gamma} a)$; we could not take small enough
$m_{\gamma}$ due to smallness of lattices. Apart from this, we
made several important observations. First, we observed a clear linear dependence
in $1/t$ for $g(t)$ (see~Figs.~\ref{fig:gtfunc} and \ref{fig:t_ext}).
This behavior can be understood as a consequence of the suppression of
backward propagation and the suppression of finite volume corrections
(owing to $m_{\gamma} L \gg 1$). When finite volume corrections are
not suppressed enough, higher poles appear in Fourier transformed
quantities and disturb a linear $t$ dependence as discussed in
ref.~\cite{Kitano:2021ecc}}. Secondly, we observed that the discretization effect for the $g$ factor
is consistent with a linear dependence in $(ma)^2$ as theoretically expected (see
Fig.~\ref{fig:cont_limit}), 
yet the statistical precision and the $m_{\gamma} \to 0$ extrapolation need to be improved.

The photon mass can be taken
small when large lattices are
available and discretization
effects will be improved significantly. 
We estimated that the required size is
$L \sim 128$. 
We would like to update our results
using larger lattice in the future.

\section*{Acknowledgements}
The authors would like to thank Masashi Hayakawa for 
discussions.
The work is supported by JSPS KAKENHI Grant Numbers JP19H00689~(RK),
JP19K14711 (HT), JP21H01086~(RK) and MEXT KAKENHI
Grant Number JP18H05542 (RK, HT).

\appendix

\section{Branch cut contribution to the effective mass}
\label{sec:A}

We consider scalar QED:
\be
S=\int d^4 x \,
\lt[-\frac{1}{4} F_{\mu \nu} F_{\mu \nu} 
+\phi^*
(-D^2+m^2) \phi \rt]
\ee
with
\be
D^2=(\del_{\mu} -i e A_{\mu})(\del_{\mu} -i e A_{\mu}) .
\ee
The two-point function
of the renormalized scalar field
is given to the one-loop level by
\be
\langle \phi_R(x)
\phi_R^*(y) \rangle
=\int_p e^{i p(x-y)} \frac{1}{p^2+\overline{m}^2-\Pi(p^2)} ,
\ee
where
\be
\Pi(p^2)
=\Pi^{\rm ana}(p^2)+\Pi^{\rm cut}(p^2) ,
\ee
with
\be
\Pi^{\rm ana}(p^2)=\frac{\alpha}{4 \pi} \lt[4 p^2-3 \overline{m}^2 + (2p^2-m^2 )  \log{\lt(\frac{\mu^2}{\overline{m}^2} \rt)} \rt] ,
\ee
\be
\Pi^{\rm cut}(p^2)
=\frac{\alpha}{4 \pi}  \frac{2}{p^2}(\overline{m}^4-p^4) \log{\lt(\frac{p^2+\overline{m}^2}{\overline{m}^2}\rt)} .
\ee
$\Pi^{\rm ana}$ is analytic at $p^2=-\overline{m}^2$ and $\Pi^{\rm cut}$ has a branch cut starting at $p^2=-\overline{m}^2$.
We consider the $\overline{\rm MS}$
renormalization and
$\overline{m}$ denotes the $\overline{\rm MS}$ mass. We work at the Feynman gauge.
One can interpret the $\overline{\rm MS}$ mass as the tree-level mass in terms of perturbative QED.

The Euclidean time correlator is given by
\begin{align}
C(t)
&=\int \frac{dp_4}{2 \pi}
\lt[ \frac{1}{p^2+\overline{m}^2}
+\lt( \frac{1}{p^2+\overline{m}^2} \rt)^2 \Pi(p^2) \rt]  \non
&=\frac{1}{2 \overline{m}}
e^{-\overline{m} t}
\lt\{1+\frac{1}{2 \overline{m}^2}
\lt[(1+\overline{m}t)
\Pi^{\rm ana}(p_4=i\overline{m})
-i \overline{m} 
\frac{\partial \Pi^{\rm ana}}{\partial p_4} \bigg|_{p_4=i \overline{m}} \rt] \rt\} \non
&\quad{} 
+\int \frac{d p_4}{2 \pi}
e^{i p_4 t} \lt( \frac{1}{p^2+\overline{m}^2} \rt)^2
\Pi^{\rm cut}(p^2)  . \label{Ct}
\end{align}
We take $p=(0,0,0,p_4)$.
Let $C^{\rm ana}$ and $C^{\rm cut}$ be
the first and the second lines of the last equality of
\eqref{Ct}, respectively.
If we approximate $\Pi^{\rm cut}$ 
as
$\Pi^{\rm cut}(p^2)\simeq -\frac{\alpha}{\pi} (p^2+\overline{m}^2)
\log\lt(\frac{p^2+\overline{m}^2}{\overline{m}^2} \rt)
=:\tilde{\Pi}^{\rm cut}(p^2)$
focusing on the point $p^2=-\overline{m}^2$,
we have an approximated  
result of the second line
of eq.~\eqref{Ct}:
\begin{align}
\tilde{C}^{\rm cut}(t)
&:=\int \frac{d p_4}{2 \pi}
e^{i p_4 t} \lt( \frac{1}{p^2+\overline{m}^2} \rt)^2
\tilde{\Pi}^{\rm cut}(p^2) \non
&=\frac{\alpha}{2 \pi \overline{m}}
\lt[e^{-\overline{m}t}
(\gamma_E+\log{(\overline{m}t/2)})-\sqrt{\frac{2 \overline{m} t}{\pi} }\frac{\del K_a}{\del a}(\overline{m}t) \bigg|_{a=-1/2} \rt]  \label{Ctcuttilde} .
\end{align}
Here $K_a$ is the Bessel function of second kind.
In giving the above result,
we used
\be
\int \frac{d p_4}{2 \pi}
\frac{e^{i p_4 t}}{(p_4^2+\overline{m}^2)^{1+\epsilon}}
=\frac{1}{\Gamma(1+\epsilon)}
\frac{1}{2  \sqrt{\pi}}
2^{1/2-\epsilon} (\overline{m}/t)^{-1/2-\epsilon} K_{-1/2-\epsilon}(\overline{m} t) 
\ee
and took the derivative
with respect to $\epsilon$. 
Our numerical analysis shows that the logarithmic term in eq.~\eqref{Ctcuttilde} is dominant for $\overline{m}t \gg 1$.
We expect that the behavior of 
$\Pi^{\rm cut}$ at $p^2=-\overline{m}^2$ is the same
for composite particles because it is determined
by IR photon.

\begin{figure}[tbph]
\begin{center}
\includegraphics[width=10cm]{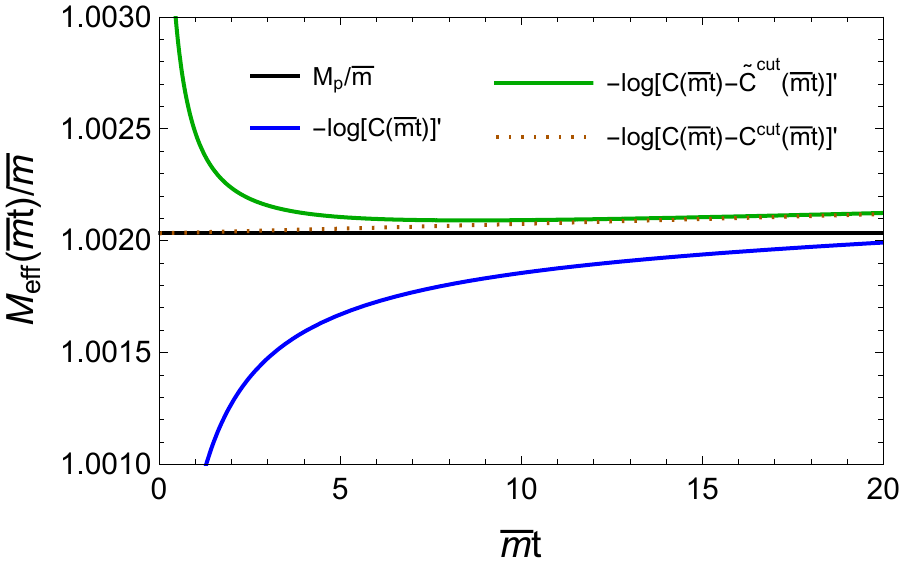}
\end{center}
\caption{Analyses of the effective mass. 
We show the straightforward analysis result (blue)
and a cut-associated contribution subtracted result (green).
We take the renormalization scale $\mu=\overline{m}$
and $\alpha=1/137$.}
\label{fig:Ct}
\end{figure}

In fig.~\ref{fig:Ct}, we give the effective mass
extracted from the Euclidean time correlator $C(t)$ (blue)
and the one from $C(t)-\tilde{C}^{\rm cut}(t)$ (green).
We give $C(t)$ numerically (by a numerical evaluation of the second line of eq.~\eqref{Ct}), while we use the analytic result~\eqref{Ctcuttilde} for $\tilde{C}^{\rm cut}(t)$.
The blue line corresponds to `lattice data' and
the green one corresponds to `corrected lattice data'.
The green line lies much closer to the exact pole mass 
value (black) than the blue line.
This would imply that the effective mass can be accurately obtained even at relatively small $t$ region 
by removing the cut effect evaluated as eq.~\eqref{Ctcuttilde} .

We make a remark regarding the figure.
If we eliminate the cut effect best, i.e. extract 
the effective mass from $C(t)-C^{\rm cut}(t)$
(where $C^{\rm cut}$ is also evaluated numerically),
we obtain the dotted line. This line has a small slope
and does not generally
agree with the pole mass value. This deviation can be 
understood as higher order [$\mathcal{O}(\alpha^2)$] effects
and thus artifact of truncated perturbation theory.

\end{document}